# Graduate Training in Quantum Information Science and Engineering: Lessons, Challenges, and a Roadmap from the NSF Research Traineeship Programs

Yohannes Abate[1], Victor Acosta[2], Alessandro Alabastri[3], Mehmet Aydeniz[4], Viktoriia E. Babicheva[5], Lincoln D. Carr[6,7,*], I-Tung Chen[8], Wandi Ding[9], Tara Drake[2], Mattias Fitzpatrick[10], Kai-Mei C. Fu[8,11], Jay Gupta[12], Kaden R. A. Hazzard[13], Sophia E. Hayes[14], Jin Hu[15], Hilary M. Hurst[16], Sohrab Ismail-Beigi[17], Ehsan Khatami[16], Junichiro Kono[3], Cheng-Yu Lai[18], Xiuling Li[19], Yingmei Liu[20], Sara Mouradian[8], Kater Murch[21], Borja Peropadre[22], Zoe Phillips[13], Daniela R. Radu[18], Akshay Sawhney[23], James Saslow[16,24], James Scoville[25], Meenakshi Singh[6,7], George Siopsis[26], David Weld[23], Chee Wei Wong[27]

** Corresponding author: lcarr@mines.edu*

[1] Dept. of Physics & Astronomy, University of Georgia
[2] Dept. of Physics & Astronomy, University of New Mexico
[3] Dept. of Electrical & Computer Engineering, Rice University
[4] Dept. of Theory and Practice in Teacher Education, College of Education, Health, and Human Sciences, University of Tennessee, Knoxville
[5] Dept. of Electrical & Computer Engineering, University of New Mexico
[6] Quantum Engineering Program, Colorado School of Mines, Golden, Colorado
[7] Dept. of Physics, Colorado School of Mines, Golden, Colorado
[8] Dept. of Electrical & Computer Engineering, University of Washington, Seattle
[9] Dept. of Mathematical Sciences, Middle Tennessee State University
[10] Thayer School of Engineering, Dartmouth College
[11] Dept. of Physics, University of Washington, Seattle
[12] Dept. of Physics, Ohio State University
[13] Dept. of Physics & Astronomy, Rice University
[14] Dept. of Chemistry, Washington University in St. Louis
[15] Dept. of Physics, University of Arkansas
[16] Dept. of Physics & Astronomy, San José State University
[17] Dept. of Applied Physics, Yale University
[18] Dept. of Mechanical and Materials Engineering, Florida International University
[19] Chandra Dept. of Electrical Engineering and Computer Science, Dept. of Chemistry, Texas Materials Institute, The University of Texas at Austin
[20] Dept. of Physics, Oklahoma State University
[21] Dept. of Electrical Engineering & Computer Science, Dept. of Physics, University of California, Berkeley
[22] IBM Quantum, IBM Research Cambridge, Cambridge, Massachusetts
[23] Dept. of Physics, University of California, Santa Barbara
[24] QuantWare, Delft, Netherlands
[25] Dept. of Physics and Meteorology, United States Air Force Academy, Colorado
[26] Dept. of Physics & Astronomy, University of Tennessee, Knoxville
[27] Dept. of Electrical & Computer Engineering, University of California, Los Angeles

## Abstract

Since 2019, eighteen NSF Research Traineeship (NRT) awards in quantum information science and engineering (QISE) and adjacent fields have been funded, constituting the largest NSF-coordinated investment in graduate QISE training in the United States. Synthesizing lessons from our programs, we work through the central tensions that every QISE graduate program must negotiate: between depth in a home discipline and breadth across the field, between structured instruction and open-ended experiential and hands-on learning, and between training individual specialists and cultivating teams that collectively cover all areas of QISE. We describe the structural and pedagogical innovations the NRT programs have developed in response, assess what is working and what remains unresolved, and sketch 12 open problems the community will need to address as QISE graduate education scales beyond the well-resourced research universities where it has up till now been mainly concentrated. Eight concrete recommendations follow: (1) adopt the startup model of team-based training as an organizing philosophy; (2) invest immediately in sensing and communication curriculum development; (3) build student agency into program governance, not just activities; (4) establish structural mechanisms for industrial engagement rather than depending on goodwill; (5) design for sustainability from year one; (6) develop graduate-level textbooks spanning all three QISE pillars: computing, sensing, and communications; (7) establish shared outcome assessment instruments across programs; and (8) develop structured mechanisms for faculty professional development in QISE.

## EXECUTIVE SUMMARY

In 2019, a Kavli Foundation workshop of ~50 quantum information science and engineering (QISE) researchers from industry and academia called for a comprehensive strategic plan for quantum education and workforce development [1]. In 2022, a community of 480 QISE researchers from across academia, government, industry, and national laboratories produced a detailed roadmap for undergraduate quantum engineering education [2]. The present paper is the graduate-level contribution to that lineage, grounded in seven years of operational experience across eighteen NSF-funded Research Traineeship programs in QISE and adjacent fields, the largest NSF-coordinated investment in graduate QISE training in the United States. These NRT programs represent parallel case studies in the design space of graduate QISE education: conducted simultaneously, over multi-year timescales, across a range of institutional types, with enough variation in approach to permit meaningful cross-case comparison. This roadmap provides the lessons from these experiments to the broader QISE community.

Against the NRT program's three stated goals (to advance cutting-edge interdisciplinary research, to produce interdisciplinary STEM professionals with technical and transferable skills for a range of careers, and to develop innovative approaches with sustainable programmatic capacity [3,4]), the quantum NRT experience provides evidence of progress on each front and an honest accounting of what remains unresolved.

The central findings of the QISE graduate education community, as represented by the eighteen quantum NRTs, are as follows. First, we find that no single structural model for graduate QISE training has proven universally superior, and that the field is best served by the deliberate coexistence of multiple program architectures rather than premature convergence on a single template. Such architectures include standalone degrees, embedded certificates, hackathons, and bootcamp-augmented research programs, and multi-institution partnerships. The right model depends on institutional context, student population, and the specific segment of the quantum workforce a program addresses. Second, across all eighteen programs, we find that the three-pillar imbalance across quantum communications, computing, and sensing in graduate QISE education represents a critical workforce gap, not merely a pedagogical one. Its core, from our programs' experience, is the severe underrepresentation of quantum sensing: educators across the NRT community independently converge on quantum computing as the curriculum default, with sensing appearing as an intended future addition rather than a current component. Quantum communication is a related but distinct problem: several NRT programs have built communication content, but a shared pedagogical framework that a new program can simply adopt does not yet exist. Both require targeted investment in curriculum development as a distinct activity from trainee support; the five-year NRT funding cycle, while valuable for proof-of-concept, is insufficient to produce the portable, validated curriculum artifacts that scale-up requires. Third, what has emerged most consistently across our programs is that student agency (defined as structured ownership by students of program elements that persist beyond their own tenure) is not a secondary feature of successful graduate QISE programs but a primary one, producing professional skills, community cohesion, and cross-disciplinary fluency that formal coursework alone does not reliably develop. Fourth, seven years of running these programs has produced a clear view about what the community lacks: the absence of graduate-level textbooks spanning engineering-oriented QISE across all three pillars is a material constraint on faculty adoption, encountered consistently in our own teaching and program-building experience, and developing such resources deserves priority comparable to developing courses themselves.

The remainder of this roadmap provides the evidence base and detailed analysis behind these positions. Sections II through VII describe the central pedagogical challenges of graduate QISE training and the structural and curricular innovations the NRT programs have developed in response. Section VIII catalogs twelve open problems the community is currently working to address: the three-pillar imbalance, quantum communication as a missing pedagogical middle, open quantum systems as an invisible substrate, bridging-role preparation, physical versus virtual training outcomes, implementing deliberate practice at the graduate level, the assessment gap, the scaling problem, formalism entry-point effectiveness, sustainability, citizenship restrictions, and the AI inflection. Section IX translates the NRT experience into eight specific, actionable recommendations for programs being designed or expanded now.

## I. INTRODUCTION

Quantum information science and engineering has arrived at an unusual moment in the history of a technical field. The theoretical foundations are mature enough to support active industrial development: companies and national laboratories are building and operating devices with increasing qubit counts, coherence times, state preparation-readout-control fidelities, sensitivity and transduction efficiency. Yet the field remains sufficiently open that a graduate student entering today can expect to encounter unsettled questions across multiple hardware platforms, algorithms in contrasting paradigms, device physics, error correction, and even basic quantum theory within a Master's or the first two years of a PhD. This has considerable workforce implications. The Quantum Economic Development Consortium formed under the National Quantum Initiative has consistently reported [5] a shortage of qualified applicants relative to open positions, with some estimates placing the ratio of open positions to qualified candidates at roughly three to one [6]. A 2024 workforce roadmap from the CUbit initiative at the University of Colorado characterized demand as outpacing supply at every credential level, from bachelor's to PhD [7].

QISE graduate education is constrained by a structural mismatch: demand is accelerating while training infrastructure remains under construction, in part because of the field's highly multidisciplinary nature. This is not the first time a field finds itself in this situation: density functional theory, semiconductor physics, laser technology, and nanoscience all passed through similar inflection points where the community needed to build training infrastructure roughly in parallel with the research itself. What distinguishes QISE is the degree to which the field is multi-disciplinary. For example, understanding a superconducting qubit processor requires lithographic fabrication, cryogenic engineering, microwave measurement techniques, quantum error correction theory, and pulse-level control software, among other skills. The development of emerging qubit platforms is increasingly inseparable from materials engineering, including crystal growth, exfoliation of layered materials, and the structural and functional characterization needed to identify and optimize quantum phases relevant for next-generation devices. Candidate topological qubit platforms illustrate this convergence, since progress depends not only on device design and control, but also on the ability to synthesize and evaluate new materials that may host protected quantum states. No single doctoral program has historically provided all of these at once, and it is a serious challenge for a PhD student to master all of them individually. A quantum sensor based on nitrogen vacancy centers in diamond requires a different but overlapping skill set: spin physics, optical pumping and readout, signal processing, noise analysis from a quantum perspective, and applications knowledge in whatever domain the sensor is being deployed. A quantum network node requires yet another combination of skills. The search for new layered and exfoliable materials platforms has similar issues, including candidate topological systems that may support Majorana-based qubits, where progress depends on the control of crystal growth, structural and functional characterization, device integration, and quantum measurement. A unifying pedagogical principle runs through the NRT experience and through this roadmap: graduate QISE training succeeds when it develops students' capacity to problem-solve, not when it just maximizes coverage of topics. We return to this framing explicitly in Section VIII.F in terms of deliberate practice theory in education.

Quantitative documentation of this landscape has begun to emerge. A comprehensive analysis of course catalogs from 1,456 US institutions, encompassing all research-intensive universities, all minority-serving institutions, and all institutions with relevant Accreditation Board for Engineering and Technology (ABET) accreditation, identified 61 institutions offering 89 distinct QISE programs [8]. The concentration, however, is stark: 54 of the 61 are PhD-granting institutions, and 28 states have no QISE program at any level. Across all 1,456 institutions, the study identified 8,456 courses whose title or description mentions “quantum”. A companion study scoped specifically to engineering departments found only 14 QISE programs housed exclusively in engineering, and documented a sharp Carnegie-classification gradient: PhD-granting institutions average 2.74 quantum-related engineering courses and 0.56 courses centered specifically around QISE topics, while master's-granting institutions average 0.31 and 0.07 respectively, and two-year and four-year institutions are effectively at zero [9]. Geographic and institutional-type gaps are highly correlated: states without QISE programs are overwhelmingly states without doctoral-level engineering research programs in this area.

The mechanical engineering pipeline deserves particular attention. It is the largest engineering discipline by bachelor's degrees granted in the US, yet the landscape study [9] identifies only approximately 100 quantum-related courses across the entire national sample of mechanical engineering curricula, and essentially one course centered on QISE. The single largest producer of engineers in the country is almost entirely disconnected from quantum training, a policy-level disconnect with direct workforce implications: dilution refrigerator engineering, cryogenic system integration, vacuum systems, and precision mechanical design are components of the quantum supply chain that mechanical engineers are best positioned to contribute to, and a pipeline that produces them without quantum literacy leaves the field systematically shorthanded.

A parallel study of upper-division quantum mechanics instruction found substantial variation in content and pedagogical approach [10]. A global survey identified 86 master's programs in quantum technology as of 2024, the majority established after 2021, with a pronounced trend toward interdisciplinary joint-faculty governance and industry internship requirements [11]. The Piña et al. data [8] quantitatively demonstrate a noticeable imbalance between the three pillars of QISE: computing, communication, and sensing. Throughout this roadmap, quantum sensing refers to the use of quantum coherence and entanglement in physical systems (such as nitrogen-vacancy centers in diamond, atomic clocks, and quantum-limited interferometers) to measure physical quantities with sensitivities unattainable by classical means; it encompasses the associated hardware, control, and readout techniques. Of the dedicated QISE courses identified across all institutions, 418 of 514 categorized QISE courses focus on quantum computing and information (81 percent); 70 courses cover quantum technology implementation (which subsumes sensing hardware) and all other QISE categories have fewer than 15 courses each. Quantum communication courses are present only as a subset of cryptography and quantum information offerings.

Piña et al. identify only 18 QISE courses nationwide with a laboratory component, 13 of them in physics departments [8]. The catalog-based method undercounts graduate-level experimental QISE content to some degree, since lab experiences housed within research groups, bootcamps, or modules within existing departmental labs are not reliably captured in catalog surveys; nonetheless, the scarcity of named experimental QISE courses against an industrial landscape that consistently identifies experimental skills as the primary hiring criterion is a serious issue. The question of what graduate students actually need to know to succeed in this workforce has been studied directly through industry interviews. Experimental skills, broadly defined to include hardware characterization, cryogenic measurement, clean-room techniques, and quantum hardware debugging, emerged as the central unifying requirement for non-PhD roles, cutting across hardware, software, and business functions [12]. A companion study found that the level of quantum expertise required varied substantially by role, from quantum-aware generalists to quantum-expert specialists, with the large middle range of quantum-proficient and quantum-conversant roles currently underserved by the educational pipeline [13]. Neither study found that the computing-communication-sensing breakdown of industry roles maps cleanly onto the computing-dominated educational programs, in many cases lacking a strong hands-on component (especially outside the QISE NRTs).

We use the workforce role taxonomy established in Asfaw et al. [2] as an organizing framework throughout this roadmap. *Quantum-aware engineers* form the broad base: STEM professionals who understand enough quantum context to participate in quantum-adjacent projects, communicate with quantum specialists, and recognize quantum-relevant problems when they encounter them. Above them are *quantum-proficient engineers*: professionals who can work directly on quantum systems, operate quantum hardware, implement quantum algorithms, and contribute meaningfully to quantum technology development without necessarily holding a research PhD. PhD-level scientists and engineers who can advance the state of the art, design new quantum systems, and drive the fundamental and applied research that underpins the field constitute the *quantum-expert* tier.

Industry interview data suggest this three-tier framework may benefit from a finer-grained intermediate tier: the CUbit workforce roadmap [7] and recent interview-based studies have identified a *quantum-conversant* category, comprising professionals who can engage credibly with quantum concepts and interface with quantum-proficient colleagues without themselves performing quantum research. This is increasingly important in public-facing, business development, and technical sales roles [13,7]. Here, we retain the three-tier structure for clarity across sections. Bridging roles, documented empirically in [14] and discussed in Section VIII.D, are distinct: they require quantum-proficient depth in at least one technical domain combined with the cross-domain fluency the conversant tier describes. This combination does not map cleanly onto any single tier. The NRT programs described herein primarily target the quantum-proficient and quantum-expert levels, though several have built explicit on-ramps for quantum-aware students from adjacent disciplines. The workforce gap documented in the QED-C workforce assessment and the CUbit roadmap is concentrated in the quantum-proficient tier (the middle of this spectrum), where the supply of graduates is most significantly below projected demand [15]. The recommendations of Ref. [15] are addressed primarily to programs trying to fill that tier, with the bridging-role design challenge representing an important open problem that the NRT experience has identified but not yet resolved.

Against this backdrop, the NRT program has been generating parallel experience from the inside. Dedicated to shaping and supporting highly effective training of STEM graduate students in high-priority interdisciplinary or convergent research areas, the NRT program operates through comprehensive traineeship models that are innovative, evidence-based, and aligned with changing workforce and research needs. Its three stated goals are (i) to catalyze and advance cutting-edge interdisciplinary or convergent research in high-priority areas; (ii) to increase the capacity of U.S. graduate programs to produce cohorts with broad participation of interdisciplinary STEM professionals with technical and transferable professional skills for a range of research and research-related careers within and outside academia; and (iii) to develop innovative approaches and knowledge that will promote transformative improvements in graduate education, with creation of sustainable programmatic capacity as an expected outcome [3,4].

Since 2019, eighteen programs in QISE and adjacent fields have been funded under the NRT program, reflecting the field's standing as one of the most heavily represented thematic clusters in the NRT portfolio. Each program is required to develop training that crosses disciplinary boundaries, integrates professional development alongside research training, and demonstrates potential for impact beyond the funding institution. NRT programs are explicitly not standalone research grants: the training model itself is the subject of inquiry, and the program expects grantees to contribute to knowledge about what works in graduate STEM education. Its predecessor, the Integrative Graduate Education and Research Traineeship (IGERT), funded roughly 125 projects over a decade but was found to lack a non-trainee mandate and rarely produced institutionalization plans, with activities typically ceasing when funding ended; the NRT program was redesigned explicitly to address both shortcomings [16]. Figure 1 places the NRT portfolio within the broader national landscape of graduate QISE credentials that have emerged since 2013.

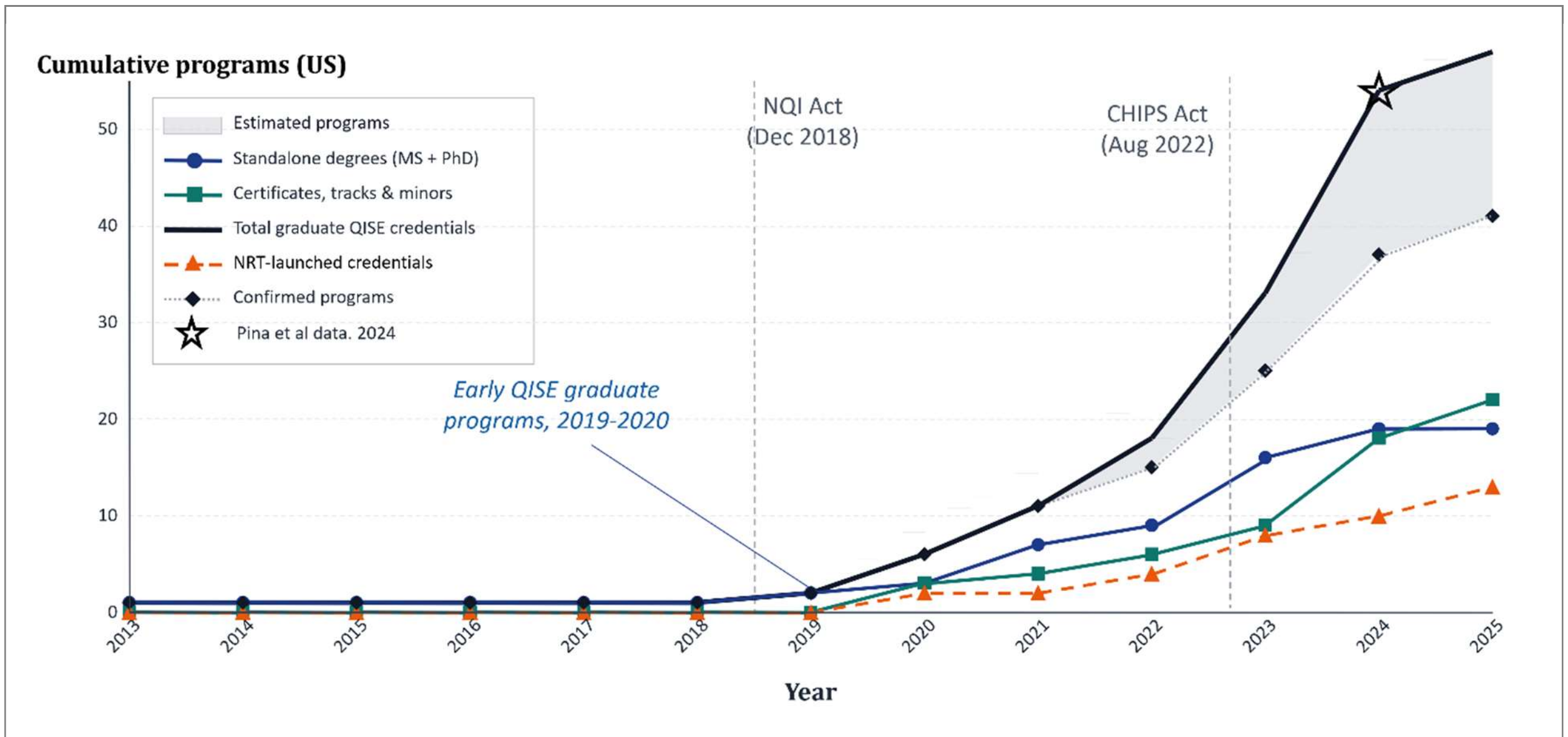

**Fig. 1.** Cumulative growth of graduate QISE programs in the United States, 2013–2025. Standalone degrees (MS and PhD) and certificates, tracks, or minors are shown separately; the total counts both. NRT-launched credentials form a significant subset. The grey band represents the approximately 54 programs identified in Piña et al.'s 2025 systematic survey [8] whose exact launch years could not be independently verified; their temporal distribution within 2020–2024 is interpolated. The National Quantum Initiative Act (December 2018) marks an inflection after which graduate program creation accelerates.

We build on two prior community syntheses that established the foundations on which the NRT programs were designed. A Kavli Foundation workshop at UCLA in 2019 [1] reported on eighteen post-secondary QISE programs and laid out the case for a common language, convergent curriculum design, and the spectrum of workforce roles from quantum-aware to quantum-expert, a vocabulary the NRT programs have operationalized. An undergraduate-focused roadmap based on a 2020 NSF workshop at the Colorado School of Mines provided a curriculum framework for entry-level QISE instruction accessible to all STEM disciplines [2]. A more recent NSF-sponsored workshop examined the case for a national center for quantum education and identified remaining infrastructure gaps at the K-12 and undergraduate levels [17]. We extend that lineage to the graduate level, with the NRT programs as the primary empirical base.

As a group, the quantum NRTs are not a representative sample of graduate QISE education broadly; they are better resourced, more deliberately structured, and more intentionally convergent than the median QISE graduate program. But this is exactly what makes them informative: they represent a series of parallel case studies in the design space of graduate QISE training, conducted simultaneously, over multi-year timescales, with enough variation in institutional context and programmatic approach to permit meaningful cross-case comparison. Throughout this roadmap we distinguish, where it matters, between findings drawn specifically from the NRT experience (practitioner observations documented across our eighteen programs) and findings drawn from the broader QISE field (landscape studies, educator interviews, and workforce analyses cited at each point). The former carry the weight of multi-year operational experience at a set of research-intensive institutions; the latter describe the field conditions within which the NRT experiments are being run.

Our analysis draws on presentations, structured discussions, and curriculum documentation from a two-day satellite meeting to the full NSF NRT Annual Meeting, held in October 2024, involving the fourteen quantum NRTs funded at that point; four have been funded since (and have contributed more recently to this roadmap). Participants included principal investigators, program coordinators, and student representatives from all fourteen quantum NRTs present at that meeting, alongside contributors from IBM Quantum, the U.S. Air Force Academy, and the Office of Science and Technology Policy as representatives of industry and federal government. Four sessions of program presentations, three sessions of student perspectives, and a dedicated half-day writing session in which participants drafted the sections of this roadmap structured the meeting. Three organizing questions define its scope. (1) What are the best practices emerging from the NRT programs? (2) How can these findings be extended to a broader engineering and interdisciplinary audience? (3) What collective lessons are ready to be shared with the community? Our goal is not to report engineering education research, although many of us do engage in these activities. For example, we do not present pre/post outcome data or systematic psychometric analysis. Rather, our goal is to synthesize practitioner knowledge, grounded in seven years of NRT program experience across eighteen programs, into a roadmap that is useful to programs now being designed or expanded across the quantum ecosystem.

## II. THE CENTRAL TENSION: DEPTH, BREADTH, AND THE FOXES-AND-HEDGEHOGS PROBLEM

Three tensions run through every QISE graduate program and recur throughout this roadmap: depth versus breadth in individual training, structured instruction versus open-ended experiential learning, and cultivation of individual specialists versus training that produces functional teams. This section addresses the depth-versus-breadth tension, which is the one program designers confront earliest and most explicitly; the experiential-versus-structured tension surfaces in Section V, and the individual-versus-team tension reappears immediately below in the startup-model discussion and throughout the practitioner evidence in Sections IV through VII.

Every QISE graduate program eventually confronts the same basic question in some form: are we trying to produce experts who have deep knowledge in one subject and can communicate across the boundaries of a team, or are we trying to produce true generalists who are fluent enough across disciplines to move through the full quantum stack on their own? As a loose description of research style, we say there are hedgehogs, who focus on one subject intensely and become world experts, and foxes, who leap from topic to topic, curious about everything. QISE graduate education is, in effect, a sustained argument about which of these styles the field needs more.

Depth-first training has the stronger institutional argument. Graduate education has always been primarily about producing expertise, and disciplinary depth is not only what advisors are equipped to transmit but also what employers, including industrial employers, have told us they value when they describe senior technical roles. A program that attempts to make every graduate student fluent across hardware, algorithms, and software at research depth is likely to produce students who are less capable at all three. The analogy to the semiconductor industry is instructive: that field supports many thousands of engineers who share a common conceptual foundation but hold deep expertise in narrow regions (e.g., process integration, device physics, circuit design, and testing and characterization) without each person having to master the others' domains. There is no reason to think the quantum industry will require a different structure at scale.

Yet breadth has its own force, particularly at this early stage of a technology field. When no one has yet worked out which problems are important and which approaches will survive contact with hardware reality, moving fluidly across domain boundaries is a competitive advantage, for research, for startups, and even early-stage industrial programs in large companies where team sizes are nevertheless quite small. Across quantum science and engineering, the field spans a broad spectrum of expertise, from quantum algorithms and software, to enabling infrastructure such as cryogenic and control systems, to materials platforms underpinning emerging qubit technologies, including photonic architectures, spin-based qubits, and vacancy-defect systems. These domains operate with distinct conceptual frameworks and technical priorities, and the connections between them remain only partially integrated. This large ecosystem, while a defining strength, raises a *quo vadis* question for the field: where should efforts be directed when it remains unclear which technological approaches will ultimately prevail? The absence of convergence around a dominant platform has a direct consequence for curriculum design: programs cannot build training around a single hardware paradigm and trust it will remain relevant, and must instead develop in students the adaptive capacity to reason across shifting platforms.

One of the student speakers at our meeting characterized herself as a natural generalist, someone who had found in quantum science the rare field that required her simultaneously to do engineering, mathematics, and software. That student was articulating something real about what this moment in the field offers and demands.

Several NRT programs have attempted to thread this needle with what might be called the startup model of graduate training. In a technology startup, the relevant unit is not the individual expert but the team: a collection of specialists who coordinate around a shared goal through a common language rather than duplicated expertise. In this model, the goal of a graduate training program is not to produce generalists but to produce specialists who can function as team members: people who know their own domain in depth and can communicate across it effectively. The question then shifts from how much quantum computing does a materials scientist need to know, to what shared vocabulary and conceptual fluency enables a materials scientist and a quantum algorithms expert to work productively together on the same problem.

This reframing is useful because it separates two things that the depth-versus-breadth debate tends to conflate: the question of what any individual student needs to know, and the question of what any given research team needs to be able to do collectively. It also points toward a different design criterion for graduate programs: in addition to asking what knowledge a student graduates with, ask what collaborative configurations a student can participate in effectively. A student who has been trained in a startup-model NRT has probably not mastered cryogenic engineering, quantum algorithms, and software verification, but she has hopefully mastered something! Importantly, she has also worked closely with people who have complementary skills, knows when to ask for help, understands enough of the answer to integrate it into her own work, and has the common language to do so without a translator. Expert educator interviews offer a clarifying analogy from software engineering: application developers do not need to understand CPU microarchitecture to write effective code, just as quantum algorithm developers may not need to master gate physics; the question is what stable abstractions exist at each layer of the quantum stack, and whether graduate programs are training students to work at the right abstraction level for their intended roles [18].

Unfortunately, the tension does not disappear on this rather experimental model. It reappears in the question of how much common language is enough, and what the appropriate vehicle for developing it is. It also reappears in sharper form when we consider which subfields of QISE a student is expected to develop even minimal fluency in. A student trained primarily in quantum computing algorithms may have very little understanding of the physics of quantum sensing: the role of spin coherence times, the connection between the Hamiltonian parameter being estimated and the quantum Fisher information, and the engineering tradeoffs between sensitivity and bandwidth. Yet sensing applications represent a substantial share of near-term quantum employment. Whether this is the employer's problem to solve in follow-up training or the university program's problem to solve at the get-go is exactly the kind of question this roadmap seeks to address.

Recent industry data make the startup model more concrete and add a complication. A systematic analysis of over 80 positions across 23 quantum companies has identified a distinct category of bridging roles, defined as positions whose main function is connecting hardware and software teams, or connecting quantum expertise to application domains, within an organization [14]. These include quantum computer operators, device and system hardware computational scientists, quantum software application developers and trainers, and quantum technology end users at companies outside the QISE core. The existence of this category as a named job family, documented empirically from employer interviews, is notable: industry has institutionalized the translation function between depth and breadth that the startup model addresses. But bridging roles require a particular preparation that is neither pure depth nor pure breadth; it requires practitioners who are fluent enough in multiple domains to recognize when translation is needed and to execute it. This is not identical to the cross-disciplinary vocabulary-building that a shared bootcamp provides. Programs aiming to produce bridging-capable graduates may need to structure experiences specifically around the translation function, not just the shared vocabulary. A complementary projection, from interviews with twenty quantum industry managers, anticipates growing demand for application-focused professionals who can bring quantum knowledge into adjacent sectors, including chemistry, finance, logistics, and AI, as the industry matures beyond hardware development toward deployment at scale [19]. One manager captured the gap precisely: there is not yet even a name for the combination of quantum knowledge and domain-area fluency that this role requires. Graduate programs that take this projection seriously face a design question the NRT experience does not yet resolve: how to prepare students for roles that are not yet fully defined. Figure 2 summarizes the relationship among the workforce taxonomy, the startup model, and the bridging-role design problem, locating the NRT programs and the unresolved bridging-role question within a single framework.

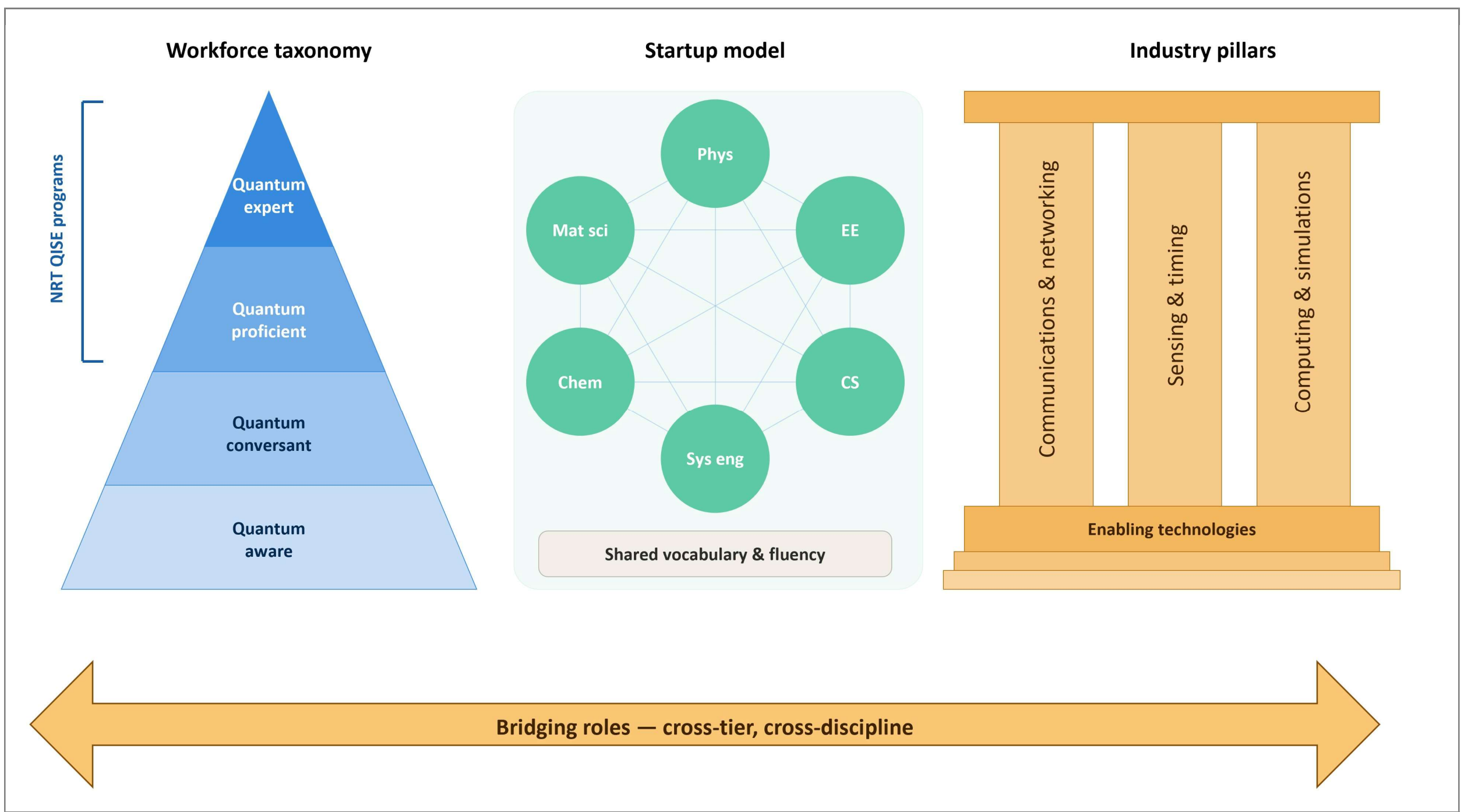

**Fig. 2.** *The workforce taxonomy, the startup model, and the bridging-role design problem.* NRT programs primarily target the quantum-proficient and quantum-expert tiers, where the workforce supply gap is most acute. The startup model organizes training around teams of specialists bound by shared vocabulary and conceptual fluency, drawn from the disciplinary clusters shown. Bridging roles, documented empirically as a named job family [14], cut across the taxonomy: they require quantum-proficient depth in at least one domain combined with conversant-tier cross-domain fluency, and no NRT program has yet built a curriculum organized explicitly around this translation function as a central organizing principle (Section VIII.D).

## III. BUILDING COMMON LANGUAGE: APPROACHES AND THEIR TRADEOFFS

The challenge of bringing students from physics, electrical engineering, computer science, chemistry, and materials science into a common intellectual space is the first practical problem every interdisciplinary QISE program faces. Students arrive with different formalisms, different intuitions about what counts as understanding, different standards of rigor, and different relationships to experimentation and computation. A physicist who has spent two years thinking carefully about the density matrix may have no intuition for what a microwave engineer means by insertion loss. A computer scientist who is fluent in circuit complexity may find the insistence on physical implementation details both unfamiliar and faintly irrelevant. An electrical engineer trained in RF systems may grasp Ramsey spectroscopy intuitively once she recognizes it as a form of phase-sensitive detection, but few programs have yet systematically built the pedagogical bridge between those two framings. A materials scientist who grows noncentrosymmetric crystals may understand symmetry breaking, domain formation, and nonlinear response at a fundamental level, yet have little intuition for how those properties translate into qubit coherence, microwave loss, or device-level performance in a quantum circuit.

Programs have addressed this through several structural mechanisms. The most common is a two-phase approach: a catch-up phase in which students from different backgrounds take tailored coursework or bootcamps to establish baseline knowledge in areas outside their training, followed by a common core taken by the full cohort as a group. The catch-up phase is necessary because graduate students arrive with varying preparation, even within a single discipline. Even within physics the difference in quantum mechanics exposure between a student from a top-ten physics program and a student from a primarily undergraduate institution can be substantial. Students from engineering and computer science frequently have little or no prior exposure to the Hilbert space formalism. The common core is where community gets built.

Several programs have found that the content of the common core matters less than the experience of taking it together. When a physicist, an electrical engineer, and a computer scientist are in the same room working through the same problems, they discover not only the shared vocabulary but the shared confusion, and the act of explaining to one another what their respective training makes obvious is itself a form of language development. The NRT at UNM developed a one year platform that explicitly takes this structure: it begins with topic-agnostic principles (harmonic oscillators, resonances, wave propagation, and two-level systems) that appear across multiple disciplines with different names and conventions, and then traces how those principles manifest in specific physical platforms. The pedagogical intention is that students arrive in the second half of the curriculum already holding a set of shared conceptual anchors that make the platform-specific material legible across disciplinary boundaries.

The WashU NRT took a different approach: rather than a standalone course, they ran low-disruption week-long bootcamps, forming multi-disciplinary teams to conduct multiple hands-on labs, coupled with lunch-hour lectures, coupled with discussions on specific quantum "keywords" as a framing device. The first bootcamp asked students and faculty together to define terms (e.g. coherence, entanglement, relaxation, polarization) and discovered, instructively, that even the faculty from different disciplines could not agree on definitions. This forced reckoning with definitional ambiguity is itself a form of common language development, and the moment of collective disorientation appears to have functioned as a kind of equalizer across disciplinary backgrounds. WashU used a quantum diamond (nitrogen-vacancy center) microscope as the common object of study, deliberately recruiting students from earth sciences, chemistry, and biology alongside the physicists and engineers, and asking the full group to brainstorm use cases for the instrument in their own fields. (Trainees learn that they can embark on a 6- to 12-month internship in a different quantum-related discipline, because the bootcamp establishes the supportive community, while illustrating some of this ambiguity among practitioners.) The choice of this particular instrument is worth noting: the nitrogen vacancy (NV) center magnetometer is one of the few quantum devices whose operating principle is simple enough to grasp in an afternoon, whose performance is close enough to the Heisenberg limit to make the quantum advantage concrete, and whose potential applications span enough fields to give students from very different backgrounds an immediate sense of relevance. The NV center platform has emerged independently at multiple institutions as a preferred common-language teaching instrument.

Rice's NRT addressed the common-language problem through a hybrid administrative–pedagogical strategy, embedding its traineeship within an existing interdepartmental graduate program in Applied Physics. While the traineeship was embedded within an existing interdepartmental Applied Physics graduate program, allowing NRT courses to be integrated without increasing student load, this structure was complemented by explicit mechanisms designed to foster cross-disciplinary fluency. These included co-advised, multi-PI research projects, as well as targeted modifications of existing courses to broaden their accessibility to students with heterogeneous backgrounds. In parallel, structured seminar participation, communication training -including 3-minute research pitches, followed by professional feedback from professionals and faculty- and team-based activities were used to reinforce a shared vocabulary and conceptual grounding.

The tradeoff is partial dependence on an existing interdisciplinary scaffold; however, several elements (such as co-mentored projects and course adaptation) are transferable to institutions without a formal interdepartmental program.

A recurring theme in the discussions was the importance of what is not taught explicitly. The common language that matters for collaborative research is not primarily a vocabulary list; it is a set of intuitions about what the central questions are, what the relevant scales and regimes are, and what counts as a good answer. This kind of tacit knowledge is harder to transmit in a classroom than in a research environment, and several programs found that the most effective common language development happened not in courses but in joint group meetings, shared seminars, and especially in the startup-model laboratory experiences described in the next section.

The Interagency Working Group on Workforce, Industry, and Infrastructure identified key concepts for QISE learners in 2020, and the NRT experience broadly confirms the value of such a framework at the K-12 level while indicating that graduate training requires something more flexible. QISE is a rapidly evolving field; a fixed vocabulary list developed in 2020 is already missing terminology that has become standard in the error correction literature since then. Programs that have succeeded in building a durable common language have generally done so by developing in students a habit of terminological vigilance, meaning an awareness that the same concept may appear under different names in different communities and that clarifying vocabulary is a first step in any interdisciplinary collaboration, rather than by transmitting a fixed set of definitions. Expert survey data confirm that this vigilance must extend to apparently foundational questions: the definition of quantum computer itself remains contested among specialists, with expert opinion spanning a flexible operational spectrum from current NISQ devices to fault-tolerant Shor-scale machines. No consensus has emerged on which qubit architecture will prove dominant [20]. The latter finding has direct curriculum consequences: programs that build hardware-specific intuition too early may need to rebuild student mental models as the technological landscape shifts. Against this backdrop of instability, a narrower convergence deserves recognition: the terms qubit and measurement basis have begun to function as architecture-independent anchors, constituting the minimum shared vocabulary that permits cross-disciplinary conversation without assuming a physical substrate. But this apparent convergence should not be over-read; at least one expert predicted that the vocabulary will continue to expand before it regularizes, a pattern familiar from early semiconductor and laser physics [18].

One specific and underappreciated axis of the common language problem concerns the entry point into quantum formalism. The dominant model, rooted in physics pedagogy, approaches quantum mechanics through wave functions, historical context, and continuous-variable systems before arriving at the two-level system and Hilbert space language that quantum computing and sensing actually require. For engineering students, this route is both slow and motivationally counterproductive. The Quantum Technologies minor at the U.S. Air Force Academy takes a deliberately different path: it begins from linear algebra and the two-state system, reaching gate-based quantum computing and initial sensing applications within a semester, without traversing the wave mechanics curriculum. The explicit design goal is to produce quantum-proficient officers (people who can reason about quantum devices, lead acquisition programs, and engage with quantum vendors) rather than physicists, and the linear-algebra-first structure reflects a specific theory about what that goal requires. The program's broad reach is demonstrable: the minor has enrolled cadets from every engineering and computer science major, not just physics concentrators. The pedagogical logic behind this choice has been confirmed at the level of individual practitioner experience: educators who have worked systematically with engineering colleagues find that standard quantum mechanics texts fail due to the wave mechanics emphasis, while texts organized around finite-dimensional Hilbert spaces bridge the gap effectively, a lesson arrived at independently across the community [21]. A related motivation drove UNM's design of two parallel disciplinary on-ramps in the first semester of the QPAQT curriculum: one for physicists and the other for chemists and engineers. These courses cover the same core material but at different levels of depth and mathematical framework. This then allows all students to study discipline-specific applications in the second half of the curriculum, allowing students from optical science, engineering, physics, and chemistry to enter a shared intellectual space from different disciplinary on-ramps rather than requiring a single common prerequisite sequence.

The common language problem also has a disciplinary imbalance dimension that deserves naming. Because QISE programs tend to be centered in physics departments and led by physics faculty, the de facto common language is often physics language, with engineering and computer science students required to do the translation work. This asymmetry is not neutral. It signals to non-physics students that they are guests in the field rather than co-owners of it, and contributes to the imposter syndrome that several student speakers identified as a barrier to participation in interdisciplinary quantum programs. Building a common language that is neutral across disciplines, rather than physics with engineering translation notes appended, is harder than it sounds, and few programs have yet achieved it consistently. One mitigation documented in practice is real-time discipline bridging: explicitly glossing the same concept in the formalisms of each represented discipline (the two-level Hamiltonian for physicists and the LC oscillator for engineers, so that non-physics students are offered a foothold in their own language rather than required to translate before they can follow) [21]. This addresses the surface symptom while leaving the structural asymmetry intact. A finding from structured educator interviews provides some grounds for optimism, though it requires careful reading: the barriers that keep non-physicists from entering QISE are, in practice, primarily perceptual rather than structural at the level of student capability: engineering and computer science students are fully capable of learning the relevant formalism when taught from the right entry point, and the perception that QISE demands a physics background is itself the principal obstacle that careful framing can dismantle [22]. The asymmetry in how programs are organized persists at the institutional level, but its main harm operates through perception; addressing the institutional structure would therefore remove a barrier that is currently holding capable students back.

There is a specific formalism gap in engineering QISE courses that the language problem partly explains. An analysis of engineering department course catalogs found that while quantum algorithms and quantum computing are the most commonly covered topics, density matrix representations and quantum measurement are among the least [9]. This is consequential because the density matrix and the measurement postulate are exactly the mathematical structures that connect all three pillars: sensing operates through measurement backaction, open quantum systems are described via the density matrix, and quantum error correction is fundamentally about measurement and feedback. The omission carries a specific pedagogical cost that independent expert data document: the dominant physics pedagogy, organized around measurement of physical observables and their Hermitian operators, does not develop the information-theoretic framing that QISE requires, specifically the shift from asking what eigenvalue does this operator return to what information does this basis choice extract. This reorientation from eigenvalue-centric to basis-centric thinking is not terminological, but structural [18]. By structural we mean that the "Hermitian operator and eigenvalue of an observable" framing organizes the entire logical flow of standard quantum mechanics instruction, not just its vocabulary; reaching QISE-relevant topics requires a different organizing narrative. Programs that skip this formalism in the name of accessibility to engineering students may be saving weeks at the cost of leaving students unable to reason across pillar boundaries later. The common language problem runs deeper than vocabulary; it runs to the mathematical substrate that makes cross-pillar fluency possible.

## IV. CURRICULUM ARCHITECTURE AND PROGRAM STRUCTURE

The structural options for housing QISE graduate training are numerous, and different institutions have chosen different paths based on their existing departmental configurations, the disciplinary backgrounds of their faculty, and the student populations they are trying to reach. The landscape study by Piña et al. found that of the 61 US institutions offering QISE degree programs, interdisciplinary joint-department structures are now the most common form, with standalone programs and physics-based degrees following. The NRT programs span most of this range. Table 1 provides a structured overview of eighteen quantum NRT programs, organized by structural type. Structural choices are not purely pedagogical. Educator interview data document that administrative approval processes, which are substantially more burdensome for new degree programs than for new courses or concentrations within existing degrees, directly shape which structural models are feasible, with the practical consequence that programs often land on the most administratively tractable structure rather than the educationally optimal one [23]. The implication is that the diversity of structural approaches in Table 1 reflects not only different theories about what works but also different institutional constraint environments. This complicates comparisons across programs and is a reason that our first community finding (that no single structural model is universally superior) is grounded in institutional reality and not only in pedagogical evidence.

*Table 1. Eighteen NSF NRT quantum programs organized by structural type. Primary pillar indicates the dominant research and curriculum focus; key innovation identifies the program's most distinctive pedagogical contribution. ★ denotes the two awards of the jointly funded Mines–SJSU NRT-QL partnership; † denotes the two awards of the jointly funded UGA–UTK QuaNTRASE partnership.*

| Program (acronym) | Institution(s) | Structural type | Primary pillar | Key pedagogical innovation |
|---|---|---|---|---|
| NRT-QL★ | Colorado School of Mines | Multi-institution partnership | Computing + sensing | Broad participation model; hardware + software tracks; student-driven programming via Society of Quantum Engineers |
| Sensor Sci. | Dartmouth | PhD fellowship + entrepreneurship | Sensing | Commercialization curriculum (IP law, finance); QuBlitz virtual lab; device-physics-first pedagogy |
| Q-STAR | Florida International University | QISE Track Coursework across Materials Engineering, Physics, and Chemistry graduate programs | Quantum materials | 360-degree training model including defined coursework, industrial internships, Individual Development Plan, and transferable skills training |
| QISE-AI | Middle Tennessee State University | Embedded in existing interdisciplinary program | Computing + materials + AI | Relationship-rich, work-integrated training; individualized development plan |
| QuGIP | Ohio State University | Standalone interdisciplinaryMS/PhD program | All three pillars (4 tracks) | Student-focused program design includes internships, flexible course menus, student-led seminars |
| ATTAQ | Oklahoma State University | Ph.D. fellowship plus Certificates | Sensing + Materials + AI | 2-faculty (QIS+AI) co-advising; internships; experiential and |

| Program (acronym) | Institution(s) | Structural type | Primary pillar | Key pedagogical innovation |
|---|---|---|---|---|
| | | | | project-based transdisciplinary graduate training |
| BRIDGECQED | Rice University | Embedded in existing interdisciplinary PhD program | Sensing + materials | Co-advisory model (NS + ENG faculty); 2-faculty advising per trainee; >150 students |
| NRT-QL★ | San José State University | Partnership with Mines NRT-QL | Computing | MSQT program at HSI; broad participation focus |
| 2D Quantum | University of Arkansas | Embedded in materials science | Quantum materials | 2D quantum materials foundry; democratization of 2D device fabrication |
| AIF-Q | University of California, Los Angeles | 7-department campus-wide traineeship | Computing + Comms.+ Sims+ Info. Theory | 3-faculty co-advising; 4 thrust layers; student association (QCSA); new QST courses; fee-for-facility; industry-academia-national lab partnerships; |
| InTriQATe | University of California, Santa Barbara | Interdisciplinary PhD traineeship | Quantum assembly | Open-ended cool-to-1mK lab course; senior-student curriculum co-design; EQuAL seminars |
| QuaNTRASE † | University of Georgia | Multi-institution partnership | Quantum networks | Networks as integration layer; first QISE graduate program in Southeast; STEM education co-PI |
| QPAQT | University of New Mexico | Certificate embedded across existing programs | Sensing + networks | 2 parallel disciplinary on-ramps, PHYS: Quantum Optics OR CHEM/ECE: Intro to Qu Tech for Chemists & Engineers (photonics/sensing focus) |
| QuaNTRASE † | University of Tennessee, Knoxville | Multi-institution partnership | Quantum networks | Networks as integration layer; first QISE graduate program in Southeast; STEM education co-PI |
| Q-CAT | University of Texas at Austin | campus-wide traineeship: spanning engineering and physical sciences. | cross-platform integration | Q-CAT introduces a cross-platform, experiential, and translational training model that integrates quantum materials, devices, and systems into a unified workforce pipeline aligned with real-world quantum technology deployment. |
| AQET | University of Washington, Seattle | Certificate embedded across existing programs | Computing + networks | Industry fee-for-capstone model; 3 parallel disciplinary on-ramps; |
| LinQ | Washington University in St. Louis | Bootcamp-augmented research | Sensing | Weeklong bootcamp with hands-on labs, multi-disciplinary group formation, and daily |

| Program (acronym) | Institution(s) | Structural type | Primary pillar | Key pedagogical innovation |
|---|---|---|---|---|
| | | program, launching 1-year Trainee interns | | lecture/discussions ; quantum diamond microscope as shared platform; “quantum keywords” as framing |
| Quantum Materials | Yale University | Certificate-first toward PhD | Quantum materials | Science communication portfolio; certificate launched before degree; 5 required courses including Big Data/AI |

Figure 3 maps the programs along two principal axes (structural type and primary pillar emphasis) drawn from Table 1.

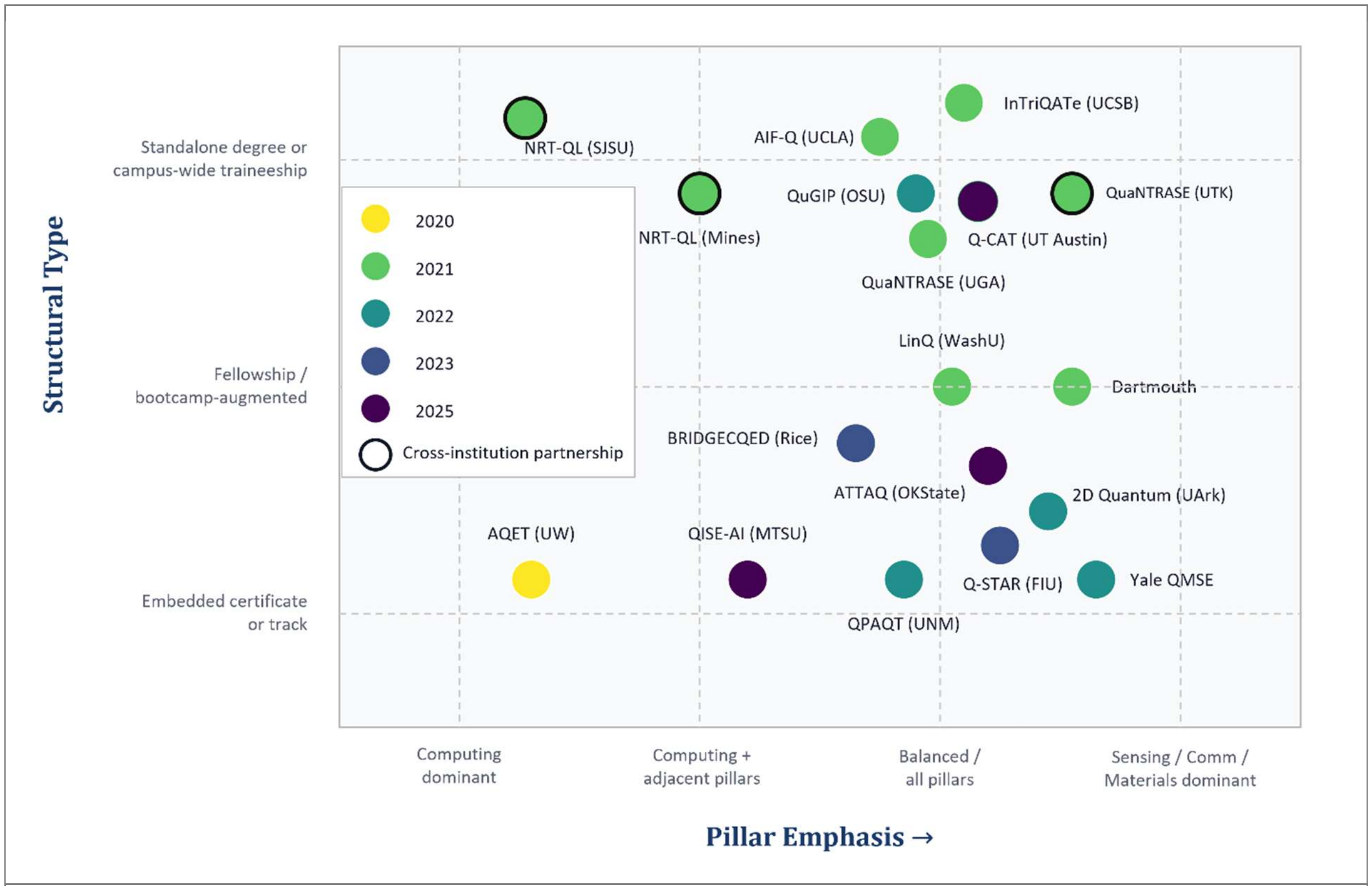

**Fig. 3.** Design space of the eighteen NSF NRT QISE programs (see Table 1 for program-level detail), plotted by structural type against primary pillar emphasis. The distribution illustrates the roadmap's first central finding: no single structural model has proven universally superior, and the deliberate coexistence of multiple architectures is itself a defining feature of the portfolio. **Bubbles with black borders denote the two collaborative multi-institution partnerships (Mines–SJSU NRT-QL and UGA–UTK QuaNTRASE).**

For example, the Ohio State QuGIP program represents one pole: a standalone MS and PhD program in QISE that spans six Departments in the Colleges of Arts and Sciences and Engineering. While providing administrative independence, this also poses challenges, since the participating units have distinct funding models, teaching-load expectations, and cultures of student support. Programs of this type require sustained institutional commitment, and may be vulnerable to changes in administrative priorities in ways that programs embedded in existing departments are not. OSU's QuGIP program features four tracks (computing, networking and communication, simulation, and materials and sensing). Students are admitted into one of these tracks and choose from a distinct menu of elective courses and research opportunities. QuGIP stands out for offering quantum communications as a primary specialization rather than a secondary application. Elective courses in this track include quantum key distribution, quantum repeater architectures, quantum teleportation protocols, and the photonic hardware (single-photon sources, quantum memories, and entanglement distribution channels) that quantum networks require. Workforce demand makes this investment concrete: quantum communications applications, including QKD systems and quantum network testbeds, represent a substantial and growing share of near-term quantum industry employment, particularly in the national security sector.

Dartmouth's NRT takes a different structural approach: a PhD innovation fellowship that deliberately ties quantum training to entrepreneurship, with coursework in law for technology, accounting, and finance, and invention disclosure alongside the technical curriculum. Its principal investigator came from IBM, and the program's design reflects a theory about what the quantum workforce needs, grounded in industrial experience rather than academic convention. The question "How will this help me get a job?" is treated not as a distraction from serious graduate training but as a legitimate and important design criterion. As part of this program, the QuBlitz virtual quantum laboratory instantiates a pedagogical approach that emphasizes quantum dynamics (specifically the time evolution of physical qubits under real pulse sequences) rather than the gate-level abstraction that dominates most introductory quantum computing curricula. Graduates who understand what a qubit actually does, this reasoning holds, are better positioned than those who understand only what gates abstractly are. A similar philosophy motivates the undergraduate textbook Quantum Mechanics for Tomorrow's Engineers [24], which foregrounds device physics and real technological systems from the outset rather than working through historical or wave-mechanical foundations. Together, these pedagogical choices represent a convergent view within the NRT community that engineering-oriented quantum instruction should be organized around what quantum systems do and how they are controlled, not around the mathematical formalism in which physicists first understood them.

The UW AQET program offers a certificate-based model that has been particularly successful in reaching students from outside physics. Roughly two-thirds of certificate participants come from engineering, computer science, chemistry, and materials science. The program uses three parallel introductory courses calibrated to different disciplinary entry points, with the explicit goal of bringing all students to a common place before the more advanced shared coursework begins. Industry partners contribute to the program through a fee in exchange for capstone project mentorship, and this fee mechanism has proven to be a workable formula for sustaining industrial engagement while maintaining the academic integrity of the capstone. The challenge of industrial buy-in, which is easy to describe but difficult to execute, is one that UW has navigated more successfully than most, though even there the program coordinators note that maintaining company engagement across multiple years requires ongoing cultivation that does not happen automatically.

UNM's Quantum Photonics & Quantum Technology (QPAQT) program shares the embedded certificate approach of UW AQET, combined with an interdisciplinary cohort model. Students have two parallel disciplinary on-ramps. In the Fall of their 2nd year in a PhD program they can take either a physics course in Quantum Optics or a chemistry/ECE course "Introduction to Quantum Technology for Chemists and Engineers" which has a photonics and sensing focus. The program culminates with all students taking a physics/chemistry/ECE cross-listed course in principles and platforms of quantum technology. In addition, students spend two semesters in a quantum photonics seminar where they develop professional skills and conduct capstone research presentations, and they engage in peer mentoring of QU-REACH undergraduate students during the summers. QPAQT has been particularly successful in engaging chemistry PhD students–of the first 50 QPAQT students, approximately 35% are from the Chemistry PhD program, ~25% are from Engineering PhD programs, and the remaining ~40% are from Physics.

Mines' NRT illustrates both the possibilities and the constraints of building a program at an institution whose student population and institutional culture differ from those of the non-polytech R1 universities where most NRT programs are housed. The program explicitly addresses broad participation through a partnership with San Jose State University, a Hispanic-Serving Institution, and has found that the presence of SJSU students in the program changes its character in ways that go beyond enrollment statistics. Hardware and software tracks (a microelectronics processing laboratory track and a low-temperature microwave measurement track alongside a quantum programming track) reflect a recognition that the experimental and computational sides of the field have different prerequisites and different career trajectories. Unlike most NRT programs, Mines has also made a deliberate decision not to center quantum computing to the exclusion of sensing and communication, reflecting a specific view about what the quantum workforce actually requires. San Jose State University was also able to launch a two-year Master's in Quantum Technology directly as a result of the NRT partnership with Mines. The Master's at SJSU, jointly managed by Electrical Engineering and Physics, was the first quantum-specific graduate program in the country at a non-PhD granting institution. A significant portion of the enrolled students are working professionals at Bay Area technology companies.

UCLA's AIF-Q (Accelerating Interdisciplinary Frontiers in Quantum Sciences and Technologies) program has a structure with four cross-cutting thrust layers (atomic qubits, photonic qubits, topological qubits, and quantum simulations) and three vertical pillars (quantum materials & light-matter interactions, microwave-optical spectroscopy, and quantum algorithms & error correction). The structure is centered on industry – academia – national laboratory partnerships, and with subsequent multi-PI team efforts. The training involves quantum computation, quantum communication networks, and quantum sensing. For sustainability and long-term impact, a hands-on quantum metrology, characterization and teaching facility is prepared, with 10-week pedagogical modules. The UCLA structure also involves dedicated space for industry – academia

– national laboratory innovations and entrepreneurship for long-term sustainability. Consistent new faculty hires address near- and long-term growth areas in quantum science, engineering and technology. In parallel, a full-time one-year Master of Quantum Science and Technology professional degree program is accredited within the University of California, with tailored coursework sequence towards industry and academic jobs and with a full-year hands-on quantum measurement-design experience. Significantly, the UCLA AIF-Q structure has teaching and design workshops, journal clubs, industry – student talks, career fairs, and hackathons led dominantly by the students and the quantum student association.

The UGA-UTK collaboration that produced QuaNTRASE deserves particular attention because it is the NRT program most fully organized around quantum networks as a research and training theme. Quantum networks are framed as the integration layer connecting quantum computing, sensing, and communication into a coherent technological ecosystem: distributed quantum resources, quantum-enhanced sensing networks, and quantum-secured communication all require the same core competencies in entanglement generation, quantum memory, and photon-photon interaction. Building graduate training around this integration layer, rather than around a single pillar, is a deliberate pedagogical and workforce strategy. It was the first QISE graduate program in the Southeast, and its geographic reach has been extended through partnerships and remote participation structures that acknowledge the concentration of existing quantum programs in a small number of coastal research centers.

The Yale QMSE certificate represents a different structural bet: launching a credential before the full degree program is ready, with five required courses covering quantum mechanics, data science, solid-state physics, and science communication, and requiring a portfolio that demonstrates both research ability and the capacity to explain quantum materials science to non-specialist audiences. This design choice (launching the certificate before the full degree) allows the program to begin enrolling students and generating cohort dynamics without waiting for the multi-year process of establishing a new doctoral degree. The science communication portfolio requirement, which has no counterpart in most QISE graduate programs, reflects a theory about workforce preparation that goes beyond technical knowledge.

The differences among these structural approaches are not merely administrative. They reflect different theories about what graduate QISE training is for: whether its primary output is individual expertise or team capability, whether it is training for research or for industry, whether depth or breadth should be the organizing criterion. There is no single right answer to these questions, and the field is probably best served by the coexistence of multiple models rather than premature convergence on a single structure. What the NRT experience does suggest is that structural choices made early (where the program is housed, what credentials it grants, how it is funded, and how students are recruited) have large downstream consequences that are difficult to undo, which is exactly why deliberate selection, with sustainability and institutional fit in mind from the outset, matters more than which specific model is chosen.

## V. EXPERIENTIAL AND HANDS-ON EDUCATION

The role of laboratory and experimental training in QISE graduate education is a point of consensus across the NRT programs, but the form it should take is not. The traditional physics dissertation model, in which a student joins a single research group, learns its techniques, and builds a dissertation around one extended project, provides deep experimental training but a narrow scope. The NRT model has pushed in the direction of broader experimental exposure, with the question being how broad, under what structure, and at what stage in the student's training.

The UCSB NRT developed a graduate laboratory course whose explicit design philosophy was to prioritize process over outcome: the sole assignment in its second year was to cool something below one millikelvin, but the pedagogical objective was not the temperature itself. The course gave students the equipment and set them free: components for a laser cooling setup, purchased in advance by a cohort of senior NRT students who had helped design the course. There was no prescribed procedure, no cookbook, and no single correct answer. Students had to brainstorm approaches, decide on a magneto-optical trap as the path forward, divide the work by expertise, make mistakes, debug equipment, and manage the logistics of a multi-person project on a ten-week timeline. UCSB's InTriQATe program frames this design deliberately as quantum assembly: the integration of disparate experimental toolsets from multiple quantum disciplines into functional composite systems. The framing positions graduate training not around mastery of a single platform but around the capacity to connect platforms together, to understand what each contributes, and to reason about the full system-level behavior that emerges from their integration. This is a substantively different intellectual objective from either deep platform specialization or broad disciplinary survey, and it has a specific workforce rationale: the most demanding near-term applications (quantum networks, quantum sensor arrays, and hybrid quantum-classical computing systems) require exactly this integration capability. That the students chose a magneto-optical trap, itself a convergent system drawing on atomic physics, laser optics, and magnetic field engineering, reflects exactly the kind of integrative thinking the course was designed to develop.

The educational theory behind this design is recognizable from the project-based learning literature, but the specific implementation carries several features worth naming. First, the senior students who helped design and equip the course received a form of training that is otherwise rare in graduate education: the full arc of experimental project management, from concept through procurement through integration and failure. Second, the open-ended nature of the assignment created conditions under which student backgrounds were assets rather than liabilities; the team's range of preparation was a feature, not a problem to be managed. Third, the absence of a prescribed answer required students to exercise a kind of judgment that problem sets do not: they had to decide not just how to proceed but what success looked like.

The challenges were significant. Equipment lead times and repair times are not well-matched to ten-week course schedules. Small-group dynamics in a situation of real uncertainty can go wrong in ways that structured coursework rarely produces. Some teams came close to the sub-millikelvin target within the ten-week window, while others did not; in every iteration, whether the apparatus would reach the target before the end of the quarter remained unresolved until the final weeks. The course was, as its designer acknowledged, an ongoing experiment.

The WashU NRT took a lighter-touch approach to hands-on training through its bootcamp structure. Rather than a dedicated laboratory course, it used short immersive experiences built around accessible quantum instruments. An optically detected magnetic resonance (ODMR) setup, an NMR demonstrator (capable of T1, T2, and chemical shift measurements)[25], and a quantum diamond microscope served as the physical anchors. The advantage of this model is its low disruption: students can maintain their regular research commitments while still getting hands-on experience with quantum systems outside their home labs, importantly working on these labs with multi-disciplinary groups (chemists, physicists, engineers, and biologists). The NV center magnetometer in particular has proven pedagogically versatile: compact enough for a bootcamp setting, affordable enough for teaching-focused institutions, and physically rich enough to illustrate coherence, decoherence, dynamical decoupling, and quantum-limited sensing all within a single platform.

The UW NRT did not build hands-on laboratory access into the curriculum, but has benefitted from QT3 - the UW's quantum technologies training and testbed laboratory. This lab is a unique user facility that performs research, develops instructional labs, and provides state-of-the-art characterization tools, which are used by groups across campus. The mission of the lab is to provide hands-on access to quantum technology hardware to accelerate both research and training in this growing field. It houses a 6-station instructional lab facility with entangled photon and NV ODMR set-ups, a research user facility with a dilution refrigerator, STM, and two single photon microscopes, and a few-qubit NV-center testbed.

The use of NV centers as a teaching platform is not accidental and deserves elaboration, because it points toward a broader pedagogical strategy for quantum sensing education. The NV center is a two-level system whose spin state can be initialized, manipulated with microwave pulses, and read out optically. This constitutes the basic toolbox of magnetic resonance, made accessible at room temperature without a large superconducting magnet. A student who builds a Ramsey sequence on an NV center magnetometer has, in the process, encountered coherence time as an engineering parameter, phase accumulation as the mechanism of sensing, and the connection between $T2^*$ and magnetic field sensitivity. These are the physical quantities that the Cramér-Rao bound and quantum Fisher information framework formalize. The formal connection must be drawn explicitly by the instructor; the lab by itself produces the intuitions that make the formal framework legible later. Whether this device-up approach produces more durable conceptual understanding than the formalism-down approach of a graduate quantum mechanics course is one of the open questions in quantum sensing education, and one that current NSF-funded curriculum development work is beginning to address empirically. The technical viability of NV center pedagogy at the undergraduate level is well established: complete coherent control sequences including T1, ODMR, Rabi oscillations, and dynamical decoupling can be implemented on benchtop hardware assembled for under $20k, with students completing individual experiments in under ten minutes [26]. The same platform scales to more advanced instruction: extending to electron-nuclear hyperfine interactions, Ramsey fringe magnetometry, and 13C coherence revivals requires only an electromagnet upgrade to the same base setup, supporting a full two-year laboratory sequence on a single hardware platform [27]. This curriculum depth on accessible, room-temperature hardware is exactly the kind of curriculum depth that a portable sensing module now in development under NSF IUSE funding aims to translate into engineering programs.

Virtual laboratory tools have emerged as a third option. The Dartmouth QuBlitz system, built on QuTiP and accessible via a web browser, allows students to simulate qubit dynamics (Rabi oscillations, pulse sequences, readout) without requiring physical hardware. The Qudit training program at Lawrence Livermore National Laboratory takes a similar approach with remotely accessible hardware. These tools address the real barrier of equipment cost and access, particularly for students whose home institutions lack quantum hardware. They also offer a specific pedagogical advantage: the ability to make mistakes at low cost, to adjust parameters and immediately observe consequences, and to develop physical intuition about quantum dynamics through interaction rather than derivation. Their limitation is equally clear: a student who has only ever simulated a qubit has not developed the experimental skills that employers consistently identify as most valued in the quantum workforce. Educator experience supports this asymmetry directly: instructors who have added physical hands-on components, such as optical QKD setups and NV center magnetometers, alongside Qiskit-based instruction report substantially better retention and comprehension, noting that physical manipulation of hardware produces qualitative conceptual shifts that software simulation does not reliably replicate [21].

Industry employer data reinforce this picture. Hands-on experimental competence (not quantum theoretical knowledge or quantum programming) was the unifying requirement across a wide range of non-PhD roles, and master's-level positions were specifically distinguished by a communication and collaboration layer: technical documentation, client-facing communication, and cross-team coordination [12]. This bachelor's/master's differentiation has direct implications for our programs. NRT graduates are primarily trained at the master's and doctoral levels. Those who lack explicit training in technical documentation and cross-team coordination are missing the specific skill that employers identify as the distinguishing credential of that degree level. Several NRT programs have addressed this directly (Yale's science communication portfolio, UCLA's capstone with industry and fee-for-facility, and UW AQET's fee-for-capstone model), but it is not universal across the community.

The Q-STAR NRT at Florida International University has embedded team-research projects in the program, assigned to teams of two-three students in each cohort, with complementary skills, up to five projects per cohort. The faculty in the research thrusts supervise the projects with assistance from trainees' mentors, typically industrial or national lab partners. This intervention takes advantage of teamwork synergy, and due to skills complementarity, encourages peer-to-peer teaching. However, this approach entails the typical challenges presented by team work, with the sequence: team forming-storming-norming and performing being inevitable.

The picture that emerges from across the NRT programs is not a single best practice but a design space. The relevant variables include the stage at which hands-on training is introduced, the degree to which it is structured versus open-ended, the role of senior students as mentors and co-designers, the balance between in-person hardware and virtual simulation, and the relationship between the laboratory experience and the student's primary dissertation research. Different programs have made different choices along each of these dimensions, and the field does not yet have the outcome data to know which choices produce better results. What the NRT experience does establish is that at least one substantive hands-on component, structured around a platform rich enough to expose coherence, decoherence, and control as engineering parameters the student manipulates directly, is more important than any single choice within the design space above. The choice of platform and pedagogical structure should be made with institutional resources and student population in mind, not in pursuit of a currently-unknown optimum.

## VI. INDUSTRY AND NATIONAL LABORATORY INTEGRATION

The relationship between academic quantum training programs and the quantum industry is an area of broadly shared aspiration and broadly documented difficulty. Every NRT program in this community expresses commitment to industry partnership; essentially, every program has discovered that the distance between expressed interest from industry and actual sustained engagement is larger than the programs initially anticipated. A complicating factor is the field's own public profile: the excitement about quantum advantage and well-publicized commercial milestones serves a valuable function in attracting students and resources, while simultaneously generating expectations that every program must actively manage; educator interviews document this as a double-edged dynamic, with hype acknowledged as both a recruitment asset and a persistent source of pedagogical friction as programs work to calibrate student expectations against technological reality [22].

Structural mismatches drive this gap, not lack of motivation. Academic programs operate on semester timescales and multi-year funding cycles, while companies operate on quarterly roadmaps and personnel changes that can end a partnership with no warning. Intellectual property agreements are a consistent friction point: university technology-transfer offices default to terms that assume all research outputs belong to the institution, while companies reasonably want some assurance that the work they are funding or mentoring will produce outcomes they can use. These frictions are not anecdotal: structured educator interviews identify IP and publication-rights concerns as the most consistently cited obstacle to sustained QISE research partnerships, while also documenting one viable structural workaround: industry directly funding student stipends with contractual protections for open publication, effectively converting the IP question into a recruiting cost rather than a research-ownership dispute [22]. The quantum industry in particular is small enough, and competitive dynamics fierce enough, that even companies with a clear commitment to workforce development may be reluctant to be seen collaborating closely with university programs that are also working with their competitors.

A risk that educator interview data raise explicitly is dependence on proprietary company platforms as course infrastructure. Programs that built coursework around a specific company's cloud quantum access (hardware simulators, proprietary SDKs, and cloud queueing systems) have found themselves stranded when companies change their access policies, discontinue an educational tier, or reprioritize away from academic support [23]. This risk is not hypothetical: several programs in this community recounted abrupt changes to commercial quantum access that forced curriculum redesigns mid-semester. Programs should treat company-provided platforms as supplements with a finite lifespan rather than as durable infrastructure, and should maintain platform-independent versions of core curriculum components. IBM Quantum's deliberate design choice, an open-access library with progress tracking built to be stable and version-controlled, reflects an institutional acknowledgment of exactly this problem. But even IBM's library is a company asset whose terms can change.

It is our experience that programs that have made industrial engagement work consistently have done so by creating structures making it easy rather than depending on goodwill. The UW capstone model, which routes industrial project proposals through an established university capstone infrastructure and charges companies a modest fee for mentorship, removes most of the friction from initial engagement. The fee is low enough that companies treat it as a recruiting cost rather than a research investment, which means that the engagement does not trigger IP concerns. The Dartmouth model of training students in the specific skills that a quantum engineer actually uses (pulse-level qubit control, hardware debugging, and realistic assessment of what quantum computers can and cannot do) represents a different kind of alignment: rather than asking companies to shape the curriculum in real time, it anticipates what the industry needs and builds it in from the beginning.

Non-profit research organizations present a model that has been underexplored in the NRT context but deserves attention. Federally funded research and development centers (FFRDCs) such as MITRE operate with longer time horizons than industrial partners, more flexibility in the IP terms they can offer, and in some cases specific program authority in the defense and national security applications that represent a large fraction of near-term quantum sensing and communication demand. Unlike a company whose quantum engagement is ultimately subordinated to commercial objectives, an FFRDC can enter a curriculum development partnership with academic objectives as the primary goal, contributing domain knowledge about real applications without the commercial constraints that make industrial partnerships fragile. This model requires further development, but the combination of academic curriculum design expertise with FFRDC application knowledge is a natural fit for quantum sensing education in particular, where the main use cases (navigation independent of GPS, magnetometry for medical imaging and materials characterization, and atomic clocks for timing and geodesy) are exactly the applications that defense and intelligence clients are funding.

National laboratory partnerships present a different profile. Labs like LLNL, NRL, and NIST share the non-commercial-partner advantages discussed above (longer time horizons than industry, more flexibility on IP) and add one that FFRDCs typically do not: access to specific quantum hardware that academic programs lack. The Mines NRT has made productive use of LLNL's Qudit remote access program and has partnered with NRL and NIST for student projects where the national laboratory scientists serve as effective co-advisors. The challenge is that national laboratory partnerships are primarily built on personal relationships, and personal relationships are not scalable. When the particular scientist at the lab who championed the partnership leaves or changes projects, the partnership typically dissolves.

IBM Quantum occupies a unique position in the ecosystem because it is simultaneously an industry partner, a hardware provider, and a curriculum developer at a scale that no academic program can match. IBM Quantum Learning, now hosted on the IBM Quantum Platform, serves roughly 500,000 users per year with a freely accessible course library that includes serious graduate-level content, including an eighteen-lesson course on quantum information theory. The existence of this resource changes the calculus for academic programs. Questions such as quantum complexity classes and quantum error correction theory that once required a specialized advanced topics course now have high-quality video instruction and problem sets available to any student who wants them, at any institution, at no cost. The implication is not that academic programs should simply point students to IBM's platform, but that the division of labor between classroom instruction and self-directed study can shift. Programs can then spend more classroom time on the things that cloud-based instruction cannot do: mentored research, physical laboratory experience, and the cross-disciplinary collaboration that requires people to be in the same room. The Qiskit Global Summer School and related IBM programs have also introduced many students to quantum computing, and program coordinators across the NRT community report that a number of these students subsequently entered NRT programs, suggesting IBM Quantum Learning functions as a de facto feeder into the graduate pipeline. IBM is one among several industry actors running similar educational initiatives at scale; others include IonQ, Quantinuum, and QuEra, and the landscape will continue to shift as more companies build workforce-facing programs.

Several programs noted a structural issue that is not primarily about partnership mechanics but about time. Incorporating professional development, internships, and industry capstones into a PhD program is additive: it extends the time to degree or compresses the time available for dissertation research. Faculty advisors, whose incentives are structured around publications and dissertation completion, are not always enthusiastic about students spending twelve weeks at a national laboratory if it delays graduation. The NRT fellowship structure addresses this in part by providing funding during the internship period, removing at least the financial argument against extended away experiences. But the cultural resistance, specifically the perception that industrial experience is a digression rather than a component of training, is harder to address through funding mechanisms alone.

## VII. STUDENT AGENCY AND COMMUNITY AS INFRASTRUCTURE

One of the most consistent findings across the NRT programs is that student-led activity is not merely a nice supplement to formal training but a structural element necessary for success. The pattern appears in too many programs and too reliably to be coincidental. A cross-program analysis of 20 NRT projects spanning non-QISE fields found that 65% of activities were open to non-trainees, with each project reaching roughly 200 additional graduate students per year beyond its funded cohort [16]. Our community's experience mirrors this pattern: the community-building function of NRT programs extends well beyond the trainee population that appears in official program counts.

At UCSB, students who completed the open-ended laboratory course in year one became the designers and builders of the course infrastructure for the following year's cohort. A senior student described this dual-training structure, in which building the course is itself a research experience, in his presentation at the workshop. At UW, a NRT trainee-led committee organized not only social events and hackathons but a public quantum lecture by Peter Shor that drew more than 700 attendees, a scale of public engagement that no faculty-organized event had achieved. A student from SJSU articulated what the quantum engineering identity looks like from a student perspective: an equal superposition of physics, electrical engineering, and computer science, with the NRT providing the structured exposure that makes the superposition real rather than nominal. At WashU, students helped set the topics for the journal club and invited the authors of papers they wanted to discuss. Outside (distinguished) speakers in quantum sensing areas were also invited to convene a “salon” prior to seminars: sharing a paper in advance of the colloquium with students, meeting solely with students to discuss the research (often turning to career advice), and building this community of NRT-affiliated scholars. (Both domestic and international students participated.) At Mines, the Quantum Engineering Society runs seminars, supports summer camps, and has effectively become a far superior professional development operation than anything the faculty could reasonably run. At UCLA, senior students mentor junior ones in an explicitly structured mentorship relationship. The UCLA quantum computing student association also drove a large number of impactful activities, from quantum design workshops (4-day ~ 300 participants per year) to career fairs, hackathons, journal clubs, and industry-national laboratory speakers events.

What these examples share is that they give students ownership of project planning and community building. Most graduate students live with sustained dependence: on an advisor for funding, on a committee for progress assessments, and on the department for teaching assignments and health insurance. Interdisciplinary programs do something different when they create spaces where students have agency and ownership. The students' energy and judgment shape outcomes that persist after they leave. And students notice this. Multiple student speakers at the satellite meeting identified this ownership dimension, not the technical content of their programs, as the most formative aspect of their NRT experience.

OSU's QuGIP program has taken this principle the furthest of any in the community, building student agency into the program's governance rather than treating it as a supplement. The program was designed with the explicit philosophy that students should help determine what they learn and how they learn it. This co-design responsibility, which could be experienced as a burden, is instead reported by students as one of the program's most valuable features. Students who have participated in curriculum design understand the tradeoffs involved in course construction in a way that students who are simply consumers of a curriculum do not, and this understanding turns out to be directly useful in industry settings where new employees are often asked to develop training materials, design internal workshops, or mentor junior colleagues. The student-driven curriculum approach also produces courses that are better calibrated to what students actually find confusing, since the people designing the courses have recently been confused by the same material.

Student agency also addresses a problem that formal curriculum design cannot easily reach: the imposter syndrome that engineering and computer science students frequently experience in physics-dominated quantum programs. Electrical engineering students who have been told, implicitly or explicitly, that they need to learn the physicists' formalism before they can contribute to the field, are unlikely to find that experience welcoming. In contrast, if EE students are instead recognized as the experts who really understand RF engineering, and are asked to teach the physicists what insertion loss means, they have a fundamentally different experience. Student-led seminars and team projects structured around complementary expertise, rather than around a common physics foundation that not everyone shares, create conditions for this second kind of experience. Programs that have invested in this kind of approach report better retention of students from non-physics backgrounds and more durable cross-disciplinary collaboration.

The question of providing food at QISE student events deserves mention, because it reflects a concrete mechanism for how communities form. Multiple programs identified catered events as a non-trivial driver of participation and community formation. Shared meals are a method for lowering the cost of showing up, and in a graduate student culture where everyone's schedule is already overcommitted, anything that reduces the cost of participation matters. The WashU bootcamps were built around lunch-hour sessions. UCSB's EQuAL seminars provided food explicitly as an incentive for attendance. The Wikipedia editing sessions at UCSB involved beer and non-alcoholic beverages. The programs that have achieved strong cohort identity have generally invested in this, and programs that have not have often found that community building was harder than anticipated.

The Wikipedia editing sessions at UCSB deserve specific attention as a model. Faculty and students gather to improve quantum-related articles on Wikipedia, and the activity functions simultaneously as a professional development exercise, a community building event, a contribution to public understanding of quantum science, and an exercise in the common language development described earlier. Students must explain quantum concepts accurately to a non-specialist audience, which requires a level of conceptual clarity that coursework alone does not always demand. It is also scalable: any program can run a Wikipedia editing session without special equipment or infrastructure. This is a non-trivial advantage in a field where most hands-on activities require expensive hardware.

The Yale and Yale-adjacent programs have made science communication a formal credential component rather than an extracurricular activity. Yale's QMSE program requires a science communication portfolio as part of the certificate. This is not pedagogical window-dressing: recent industry role taxonomy work has identified Education Advocates as a named role category within quantum companies, defined as professionals whose function is outreach, public engagement, and building quantum literacy in external audiences [14]. Programs that build science communication skills are preparing students for a career track, not just a personal development activity. A related problem deserves naming: interview data show that QISE students are broadly unaware of the range of roles that exist in the quantum industry [28]. Students cannot exercise agency toward careers they have not heard of. Programs that bring in industry speakers across role types (not only PhD scientists but also technical sales, project management, business development, and field deployment engineers) address this directly. The Dartmouth NRT, with its explicit entrepreneurship and commercialization curriculum, has taken this furthest, but the approach is adoptable in any program through speaker series, industry site visits, and career development workshops structured around role diversity rather than the standard research-career-only framing.

## VIII. OPEN PROBLEMS

The NRT experience has clarified what the central open problems in graduate QISE education are, even where it has not resolved them. The twelve problems below fall into three groups. Six (A-F) are tractable at the program level and are already subjects of active curriculum development or pedagogical experimentation. Three (G-I) require field-wide coordination — assessment infrastructure, faculty pipeline, and cross-program outcome studies of the formalism entry-point — that no individual institution can initiate alone. The remaining three (J-L) reflect external constraints on the community rather than pedagogical questions internal to it: the post-five-year sustainability question every NRT program will eventually face, citizenship restrictions on federal traineeship funding, and the AI inflection the field is navigating in real time.

### A. Three-pillar imbalance

We have known for a while that quantum computing has attracted the majority of educational QISE investment at the graduate level. However, we now have precise data, thanks to the work of our colleagues [8]. Quantum sensing and quantum communication remain nearly absent from engineering graduate programs as standalone subjects. The landscape data are reinforced by practitioner interviews: educators developing interdisciplinary graduate QISE courses independently converge on quantum computing and algorithms as the course core, with quantum sensing appearing as an intended future addition rather than a current component, a pattern that reflects demand signals and available curriculum resources, not any deliberate choice to exclude the other pillars [21]. The sensing deficit is, however, partly an organizational rather than an absolute gap: an engineering landscape study found that the majority of engineering courses explicitly mentioning quantum sensing appear not within QISE-labeled courses but within nano- and materials-science courses [9]. Sensing knowledge exists in engineering curricula; it is not being identified or scaffolded as quantum sensing. This distinction matters: closing the three-pillar gap requires not only creating new sensing courses but also helping existing courses recognize and connect their sensing content to the broader QISE framework. This imbalance is not purely a curriculum failure; it also tracks the composition of the industry itself, where quantum computing hardware companies dominate employment relative to sensing and communication companies, creating a self-reinforcing loop between workforce demand and educational supply [14]. The Barnes et al. workshop [17], which convened experts to scope a proposed national quantum education center, explicitly bounded its proposed center to K-16, concluding that graduate-level curriculum development, including sensing and communication, requires separate, sustained investment of the kind that individual NRT programs cannot provide. This roadmap's NRT community is the natural locus for initiating that investment at the graduate level.

The depth of the lack of quantum sensing education can be illustrated concretely. Although few academic programs in the US teach quantum sensing to engineering students outside of physics classes, the quantum sensing industry is actively hiring engineers with sensing-relevant skills: noise analysis, signal transduction, interferometry, and platform-specific knowledge of NV centers, atomic clocks, optomechanical sensors, and related systems. The engineering skills required are neither trivial nor inaccessible: RF engineers understand coherent detection, EE students understand noise floors, dynamic range, and sensitivity-bandwidth tradeoffs, materials scientists understand defect physics, but no one has yet built a pedagogical framework that connects these existing skills to the quantum sensing formalism in a way that is accessible to engineering graduate students and validated as effective. An NSF IUSE grant currently underway, in collaboration between Colorado School of Mines, UC Davis, and MITRE, is among the first systematic attempts to do this. The project is building a portable quantum sensing module centered on nitrogen vacancy centers in diamond, deployable across engineering programs at a range of institution types, using a relatively inexpensive mobile NV magnetometer platform. The project is investigating three research questions that are themselves open: what teaching methods in quantum sensing create more engineering student engagement and broad participation; what minimal set of key physics concepts is necessary for an engineering student to understand quantum sensing; and what affordable hardware platform can be created for deployment in teaching-focused schools. The NV center platform's established viability as a teaching instrument at room temperature and sub-$20k cost, documented in Section V [26,27], provides the technical foundation from which these research questions proceed.

These questions deserve to be recognized as research questions, not implementation questions. In particular, engineering students who encounter quantum mechanics primarily in physics courses frequently come away feeling that the quantum formalism is a little awkward and inaccessible, and that the applications are somewhat remote from engineering practice. A sensing curriculum built around a device that engineering students can calibrate, characterize, and push toward the shot-noise limits is a specific pedagogical response to this identity problem, but whether it actually works, and for which student populations, is unknown. The second question, concerning the minimum viable physics for quantum sensing, is the sensing-specific version of the depth-breadth problem. For quantum computing, the community has largely converged on linear algebra plus two-level systems as the entry point. For sensing, no such consensus exists: whether the quantum Fisher information and Cramér-Rao bound should be introduced early as the fundamental framework, or reached device-up after students have built intuition through experimental work, is an unresolved pedagogical question with real curriculum design consequences. The third question, concerning affordable hardware, is important for access as much as for pedagogy. An NV center magnetometer that can be purchased and maintained by a teaching-focused school without a quantum center changes who can offer quantum sensing education and, therefore, who can train quantum sensing engineers.

Enabling technologies (cryogenic, microwave, photonic, and materials engineering infrastructure that all three pillars depend on) sit alongside the three-pillar framing as a supporting layer of the quantum supply chain. While touched on in Section VIII.H in the context of manufacturing technicians, this layer deserves recognition as a structural curriculum problem in its own right.

### B. Quantum communication as the missing middle

The three-pillar structure of QISE (computing, sensing, and communication) is familiar, but in graduate education, the third pillar occupies an ambiguous position. It is not absent the way open quantum systems are absent: several NRT programs have quantum communication as an explicit component, and two programs (QuaNTRASE and OSU QuGIP) have made it a primary focus. What is absent is a common pedagogical framework for quantum communication that is as developed as the quantum computing curriculum. The reasons are partly historical and partly structural. The quantum computing curriculum has benefited from massive investment by industry (IBM, IonQ, Google, PsiQuantum, Quantinuum, QuEra, among others) in making their hardware and software accessible to students, and from a decade of community development of open-source tools, textbooks, and online courses. Quantum communication, by contrast, is largely a research-to-defense-to-standards pipeline in which the primary customers are national security agencies rather than consumer technology companies. The commercial deployment of QKD systems has happened primarily in Asia and Europe rather than in the US. The result is that the pedagogical ecosystem for quantum communication is thin: there are good research papers, active government testbed programs, and a growing set of NIST standards for quantum key distribution protocols, but the kind of structured introductory curriculum that a new program can simply adopt does not yet exist.

What QuaNTRASE and QuGIP have learned from building quantum communication content from scratch is directly relevant to the broader community. The core conceptual challenge is that quantum communication sits at the intersection of quantum information theory (natural to physicists) and optical engineering (natural to electrical engineers), but requires both at the same time in a way that neither community is typically prepared for. A student who understands quantum teleportation abstractly but has never thought about photon loss budgets in fiber will struggle to reason about realistic quantum repeater performance. A student who understands optical fiber engineering well but has not internalized the no-cloning theorem will not understand why quantum communication is different from classical optical communication in a way that matters. The common-language challenge is real, and the solution that has emerged in these programs, starting from physical platforms (specifically single-photon sources and quantum memory) and building up to protocols, mirrors the device-up approach that has worked for sensing education. The UCLA AIF-Q quantum communications network is a high-rate multi-node all-to-all connectivity testbed, involving industry and multi-university partners. It teaches not only the physical layer of entangled biphotons, optical-microwave qubit transduction, and various receivers, but also the network architectures, efficient error correction protocols, higher secure-key management layers, and information-theoretic aspects. UNM QPAQT program's quantum photonics work, spanning NV center spin-photon interfaces, cold-atom quantum memories, and photonic integrated circuits, exemplifies the platform-based approach: students who work with physical interfaces between spins and photons develop intuitions about fidelity, loss, and mode-matching that are directly transferable to quantum network design.

### C. Open quantum systems, an invisible substrate

Every quantum device operates in a noisy environment it cannot be perfectly isolated from. Decoherence is not a perturbation to be corrected for, it is a central engineering problem of the field. T1 and T2 are not physical constants to be looked up; they are design parameters that depend on material choices, fabrication quality, operating conditions, and control pulse design. The Lindblad master equation and quantum trajectories methods, which govern the dynamics of open quantum systems in the Markovian limit, are to quantum device engineering what the Navier-Stokes equations are to fluid mechanics: the foundational framework from which quantitative engineering proceeds. Yet they are taught, when taught at all, as a chapter in advanced quantum optics courses pitched at physics PhD students. Their engineering content is almost entirely invisible in existing curricula.

What would it mean to teach open quantum systems as an engineering subject? Perhaps it would mean treating dynamical decoupling as a control engineering problem (a sequence of pulses designed to suppress dephasing by refocusing the spin echo, analyzed in terms of its filter function in the frequency domain) rather than as a quantum error mitigation technique with a separate theoretical framework. It would mean teaching students to read a $T_1$-$T_2$ plot the way an RF engineer reads a noise figure plot: as a design constraint that interacts with other system-level specifications. It would mean understanding the Markovian approximation not as a mathematical convenience but as a claim about the bath correlation time relative to the system dynamics, with specific implications for what control strategies will and will not work. And perhaps most far-reachingly, it would mean understanding weak partial measurement and feedback beyond error correction.

A proposal currently in development by members of this group aims to build exactly this curriculum, targeting engineering graduate students who have a working knowledge of classical control theory and differential equations but little or no prior exposure to quantum mechanics at the density matrix level. Whether the prerequisite structure implied by this approach, connecting classical control directly to quantum master equations and quantum trajectories rather than traversing the standard physics sequence, actually works is itself an open research question. A partial existence proof exists at the hardware level: NV center teaching labs have demonstrated that students encounter $T_2$ coherence engineering and dynamical decoupling as measurable design parameters rather than abstract theory (see Section V and [26,27]). Whether this device-up entry into open quantum systems content transfers to graduate engineering audiences is the open pedagogical question.

**D. Bridging-role preparation**

The startup model provides a framework for team-based training, but it does not yet answer a more specific question: how do you train someone whose primary role is the translation function itself? Industry has now documented bridging roles as a named job category (quantum computer operators, device-software interface specialists, and quantum technology end users) whose defining competency is moving fluently between domains rather than specializing within one [14]. These roles do not map cleanly onto any single tier of the workforce taxonomy: they require quantum-proficient depth in at least one technical domain combined with the cross-domain fluency the conversant tier describes. Neither depth-first nor breadth-first training specifically targets this combination. The preparation required is not identical to the cross-disciplinary vocabulary-building that a shared bootcamp provides, nor is it reducible to depth in one area plus familiarity with adjacent ones. It requires practitioners who can recognize when translation is needed, understand both sides of an interface well enough to execute it, and operate productively in the ambiguous space between technical teams. No program in the NRT community has yet built a curriculum explicitly organized around the translation function. Several programs have implemented components such as cross-department co-advising, professional development seminars, and modified disciplinary courses that develop translation-related skills, but the translation function has not yet been adopted as a central organizing principle. The startup model produces students who can work in teams, with translation occurring organically; producing students who are themselves suited for bridging roles is a different design problem, and the NRT experience does not yet resolve it.

## E. Physical versus virtual training outcomes

The design space for experiential and hands-on QISE education has at least five independent variables: the stage at which hands-on training is introduced, the degree to which it is structured versus open-ended, the balance between physical hardware and virtual simulation, the role of senior students as co-designers of laboratory infrastructure, and the relationship between the laboratory experience and the student's primary dissertation research. Programs across the NRT community have explored different positions in this space: from the fully open-ended cool-to-1mK assignment, to bootcamp-structured NV center sessions, to browser-accessible qubit simulators. The field does not yet have the outcome data to say which positions produce graduates who are more capable, more broadly prepared, or better matched to the workforce roles the quantum industry is trying to fill. Employer data establish that hands-on experimental competence is the primary hiring criterion for non-PhD industry roles [12], and educator experience consistently demonstrates that physical hardware produces conceptual shifts that software simulation does not replicate [21]. But whether open-ended project-based approaches outperform structured laboratory courses for producing the specific capabilities employers identify (and what role virtual tools should play alongside physical ones) is an open research question in quantum engineering education. It is the most operationally consequential unknown in the community's collective self-knowledge about what its own programs are actually producing.

## F. Deliberate practice as a design principle for QISE graduate training

Graduate training in any field develops expertise slowly and unevenly, and the reasons are now better understood than they were a decade ago. Research on the development of scientific and engineering expertise identifies a set of 29 problem-solving decisions that define what experts actually do when they solve authentic problems across all science and engineering disciplines. These range from choosing what concepts are relevant to reflecting on whether the solution approach is working. Standard coursework, including standard graduate coursework, has been shown in undergraduate physics contexts to give students practice on only two or three of these decisions [29]. The rest are stripped out by the structure of problem sets, which tell students what method to use, what approximations to make, and what information is relevant. Typically, students learn information but not how to use it. In contrast, the deliberate practice framework, grounded in cognitive science research on memory and expertise, developed by Ericsson and colleagues, and applied to science and engineering education by Wieman and collaborators [29,30], identifies this as the central deficit: what students need is intensive, focused practice on specific decision-making subskills, with feedback on how to improve, repeated until the subskill becomes automatic. The 29-decision characterization of expert problem-solving is drawn primarily from undergraduate and professional contexts; it is an open problem as to how to apply this to QISE graduate training.

In fact, the NRT programs have already developed program elements that are consistent with deliberate practice principles: the UCSB open-ended laboratory course, WashU's bootcamp structure, OSU's student co-design model, and the startup-model team projects across many programs all require students to make decisions that standard coursework removes. But none of these were specifically designed using the 29-decision framework as an organizing criterion, and none have outcome data comparing the decision-practice density of their training against counterfactual approaches. The deliberate-practice intervention studies in graduate quantum mechanics [31] represent a first attempt to apply this framework in a QISE context, and demonstrate that fine-grained course-level assessment informed by this framework is feasible. Whether designing a full QISE graduate program explicitly around the deliberate-practice taxonomy would produce measurably better outcomes is an open question that the NRT community is well-positioned to investigate. First steps would include (i) mapping each program element to the decisions it gives students practice on; (ii) examining authentic QISE problems across topic areas to identify what specific knowledge is actually needed to make the problem-solving decisions those problems demand, a process that itself helps clarify which topics are essential and which are peripheral; (iii) identifying which decisions are chronically underrepresented; and (iv) building structure specifically to fill those gaps.

### G. The assessment gap

None of the programs represented in this community has systematic outcome data that would allow a comparison between their graduates and a counterfactual: graduates from similar programs without the NRT structure, or graduates from the same programs before the NRT launched. The SURE framework for undergraduate research experience evaluation has been proposed as a possible template, and an evaluation team from Tennessee and Georgia is working to extend a similar framework to the NRT context. But the outcomes QISE graduate training is supposed to produce (the ability to function effectively in an interdisciplinary research team and the capacity to learn across disciplinary boundaries over a career) are not easily captured by standard metrics of time-to-degree and publication count. Work on misconceptions in quantum physics instruction [32] and the deliberate-practice intervention studies in first-semester graduate quantum mechanics [31] suggest the kind of fine-grained assessment that is possible at the course level; extending this to program-level interdisciplinary outcomes is a substantially harder problem. Educator interview data confirm that the assessment gap is structural rather than individual: most programs rely on anecdotal feedback rather than systematic data collection, not because program directors are indifferent to evaluation but because the field lacks purpose-built assessment instruments for QISE at the program level [23]. A recent community workshop independently identified assessment infrastructure as a coordination challenge requiring investment beyond any individual program [17]. The problem is compounded by the fact that most NRT programs do not publish their external evaluation findings; the data collected on program efficacy is largely lost beyond the required NSF reporting system, making it impossible to build a cumulative evidence base across the portfolio [16]. This community's position statement on assessment and the practical reality documented here converge on the same prescription: systematic, cross-program assessment infrastructure is not an add-on to good program design but a precondition for knowing whether the design is working.

**H. Scaling**

The NRT programs collectively fund on the order of 100-200 trainees per year, with a broader reach through non-trainee activities documented in cross-program analyses [16]. Industry workforce assessments suggest the field will need graduate-level supply that is order-of-magnitude greater than current programs produce within a decade [15], though specific target numbers vary by projection. That demand will also need to come from a much broader range of institutional types and geographic locations than the current NRT cohort. A prerequisite for that expansion that is rarely named explicitly is linguistic: the common vocabulary of QISE is itself still stabilizing, and programs at institutions without deep faculty expertise will find it harder to track the moving terminological frontier. Expert data confirm that even well-established terms like qubit and measurement basis are best understood as provisional convergence points rather than settled definitions, and that the vocabulary governing hardware-software interfaces, noise models, and cross-pillar applications is still in active flux [18]. Curriculum materials developed for one terminological moment may require significant revision within a few years, a cost that teaching-focused institutions are least equipped to absorb. The curriculum innovations and structural experiments documented in this roadmap were developed at research-intensive universities with substantial resources for graduate training. Translating them to R2 institutions and programs outside the current NRT network is not automatic.

The scale-up problem involves not only resources but faculty: many of the most effective elements of NRT programs depend on faculty who are themselves trained in QISE, and the pipeline of QISE-trained faculty is still very small. Educator interviews confirm this is not an impression but a documented program-level constraint: nine of fifteen QISE educators interviewed across a range of institutions explicitly identified insufficient QISE faculty expertise as a weakness in their program's ability to deliver a full curriculum [23]. Institutions with dedicated quantum centers and deep faculty rosters can cover the full range of computing, sensing, and communication topics; most institutions cannot. The scale-up problem is therefore also a faculty development problem that many individual institutions cannot solve.

A parallel dimension of the scaling challenge concerns the workforce tier that the NRT programs do not directly serve. For example, industry projections anticipate growing demand for bachelor's-level manufacturing and deployment technicians as the field transitions from laboratory proof-of-concept to manufactured products including fiber splicing, optical assembly, cleanroom fabrication, and cryogenic system maintenance [19]. Rather than training specialists narrowly aligned with a single platform, there is a growing need to prepare a workforce that is adaptable, cross-disciplinary, and conversant across the full QISE stack, from materials discovery and device physics to systems engineering and quantum information. Trainees must be equipped not only with deep expertise in their home discipline, but also with the ability to translate across domains, engage with multiple technological paradigms, and navigate a landscape in which the dominant platforms have yet to be determined. Workforce development shifts from producing platform-specific experts to cultivating flexible innovators capable of contributing across evolving and potentially convergent quantum technologies.

Finally, we want to emphasize that the NRT community trains the top of a talent stack whose lower tiers remain largely unaddressed. The explicitly portable design of the sensing module under development, using a commercial-grade, relatively inexpensive NV magnetometer platform and a widely used learning management system, is a direct response to the institutional-access dimension of this scaling concern. Whether portability in hardware and software translates to adoptability in practice, across the range of institutional cultures and resource environments that define the broader US higher education landscape, is something only deployment will reveal, and certainly remains an open problem in scaling up broader QISE education.

**I. Formalism entry-point effectiveness**

Section III documents a convergent practitioner finding: the dominant wave-mechanics-first physics pedagogy fails engineering students, and a linear-algebra-first entry (beginning from two-level systems and finite-dimensional Hilbert spaces) produces broader accessibility and faster arrival at practically relevant QISE content [18,21]. This convergence is now well established in practitioner experience and educator interview data. What remains open is the harder comparative question: does the linear-algebra-first approach produce more durable conceptual understanding, and specifically the reorientation from eigenvalue-centric to basis-centric thinking that cross-pillar fluency requires? The density matrix and measurement postulate gap documented in engineering curricula [9] suggests that the dominant physics entry point produces graduates with a structural blind spot in what measurement means in quantum information contexts [18]. But whether an alternative entry point systematically closes this gap (and for which student populations) has not been demonstrated with controlled outcome data. The answer matters well beyond QISE: it is a question about what the minimal mathematical prerequisites actually are for cross-pillar quantum fluency, and its resolution would have direct curriculum design consequences for the engineering programs that enroll the largest number of students.

**J. Sustainability**

NRT grants are five years and non-renewable. The programs described in this roadmap are, with few exceptions, entirely or substantially dependent on NRT funding for their distinctive features: the bootcamps, the graduate training, the laboratory courses, the internship support, and the community-building activities. What happens after year five is a question that several programs are currently navigating and that all of them will face. Cross-program analysis of NRT projects in other fields documents the same friction sources our community has encountered: faculty engagement declining over a multi-year span, administrative structures that fail to incentivize cross-unit participation, and the difficulty of sustaining high-cost activities like symposia without dedicated post-award funding [16]. Some have found sustainability paths through industry fee-for-service models, institutional cost-sharing, and/or integration of NRT courses into regular departmental curricula. The IUSE grant mechanism, which funds curriculum development as a distinct activity from trainee support, offers one model for sustaining pedagogical innovation after a traineeship ends: a quantum sensing module currently in development under NSF IUSE funding is explicitly designed to be portable and adoptable by programs that had no involvement in its development. Whether curriculum artifacts developed under one funding mechanism can sustain adoption at institutions with very different resource levels is its own open question. Beyond curricular and programmatic continuity, sustainability in QISE-focused NRT programs can also be driven by the rapid expansion of quantum-related research tracks and the parallel emergence of materials platforms with intellectual property (IP) potential. Advances in quantum-relevant materials are positioned at the interface of fundamental discovery and translational application. These materials systems not only generate publishable knowledge but also create opportunities for IP and technology transfer, establishing alternative sustainability pathways beyond traditional funding models, while integrating trainees into the innovation pipeline. Universities with established innovation ecosystems, such as startup incubators, translational research programs, and industry partnerships (e.g., StartupFIU) can leverage these developments to support spin-offs, facilitate access to emerging technologies, and broaden community engagement with QISE advances. In this model, sustainability is achieved not solely through continued funding, but through a self-reinforcing cycle of discovery, IP generation, and workforce development.

**K. US citizenship restriction**

NSF and most other federal funding sources require that NRT fellows be US citizens or permanent residents. This is a structural constraint on the breadth of the quantum workforce that has no straightforward solution within the current policy environment, and which sits in obvious tension with the historical and current demographics of STEM graduate enrollment in the United States. The programs in this community are acutely aware of the problem, which surfaced repeatedly in discussions at the satellite meeting, but have few tools to address it beyond strategic cost-sharing arrangements that allow non-eligible students to participate in program activities without receiving NRT fellowships. Institutional matching funds are one concrete mechanism with documented pedagogical benefit: at least one NRT program in this community has secured matching support for roughly two additional fellowships per year specifically for international students, which is insufficient to match applicant demographics but substantially improves cohort composition and, in the practitioner experience of that program, strengthens the training of domestic students through the intellectual diversity it makes possible. The broader policy question of how the US will staff a quantum workforce in an environment where graduate STEM enrollment is heavily international, while federal funding restrictions limit who can receive training support, is one that this community is not positioned to resolve, but is positioned to name clearly.

**L. AI inflection**

The role of artificial intelligence in quantum science and engineering education is changing rapidly enough that any roadmap paper written in 2026 risks being superseded before it is published. AI tools for quantum circuit design, multi-qubit processor RF control pulse sequences, error correction, and materials property prediction are becoming sufficiently capable that graduate training programs need to grapple with what it means to educate students in a field where AI is an active participant in research, not just a tool for data analysis. Across our programs, large language models are already being used by students in ways that both assist and undermine learning. The specific problem of assessment integrity in a field where AI can complete most conventional problem sets is one that the community is actively working on, with no consensus solution. UCSB and UCLA's approaches of laboratory courses that specifically require physical hardware interaction (students must test or build-debug real equipment, which AI cannot do for them) is one answer. The presence of AI as a subject of research in the field itself creates new interdisciplinary bridges: machine learning approaches to quantum materials characterization and quantum control create natural connections between the quantum education community and the AI education community that did not exist five years ago. The quantum sensing curriculum being developed at Mines uses NV center magnetic imaging as a platform in part because the data analysis pipeline for NV magnetometry is a natural entry point for ML methods, creating pedagogical continuity between sensing physics and data science.

## IX. RECOMMENDATIONS AND A PATH FORWARD

The following recommendations are drawn directly from the NRT experience documented in this roadmap. They are presented as concrete prescriptions, not as aspirations, because the field has now accumulated enough operational evidence to support definite positions. Programs being designed or expanded now should consider each in the context of their specific institutional circumstances. The NRT programs are better resourced than the median graduate QISE program, and not every mechanism cited here is directly portable to teaching-focused institutions without quantum centers. Where the NRT evidence speaks most clearly to resource requirements, we note it; for recommendations whose resource demands are inherent to their goal, programs should plan accordingly.

1. Adopt the startup model of team-based training as the organizing philosophy for graduate QISE programs. The most effective NRT programs are producing specialists who can function as team members: people with real depth in a home discipline who share enough common language with adjacent specialists to work productively together on compound problems. Programs should structure their curriculum, their advising models, and their laboratory experiences explicitly around this objective, rather than either specialization alone or breadth-for-its-own-sake. Honest advising is required about what industry and national laboratory roles actually require: depth in one area, communicative fluency across several, and the judgment to know when to ask for help. Programs targeting bridging roles specifically (Section VIII.D) will need supplemental structure organized explicitly around the translation function, not only the shared vocabulary that startup-model training provides.

The rapidly evolving nature of quantum technology deployment reinforces this: rather than producing specialists narrowly aligned with a single platform, programs should cultivate graduates who are adaptable across the full QISE stack, from materials discovery and device physics to systems engineering and quantum information. Such graduates should be equipped to translate across domains and move effectively through a field in which the dominant platforms have not yet been settled.

2. Invest immediately in quantum sensing and quantum communication curriculum development, and treat this investment as distinct from trainee support. The three-pillar imbalance is the most actionable finding of this community, and sensing and communication are distinct problems within it. Quantum sensing is the more acute gap: the pedagogical ecosystem is thin, engineering programs fail to connect their existing sensing content to the quantum sensing framework, and no validated portable curriculum yet exists. Quantum communication is further along (several NRT programs have built it from scratch), but what is missing is a shared pedagogical framework that a new program can adopt without starting over. Both represent growing shares of near-term quantum industry employment. NV center magnetometer platforms, atomic clock demonstrations, and fiber-based quantum key distribution setups are now available at price points that make them viable outside major research centers. Programs should establish at minimum one laboratory experience or dedicated course module in sensing and one in communication within the first two years of operation. The NSF IUSE mechanism provides a funding pathway for curriculum development that does not require trainee slots; programs should use it. Graduate-level experimental QISE laboratory courses are part of this investment: experimental competence is what the non-PhD workforce consistently identifies as its most valued hiring criterion [12], and the pipeline that produces it is narrower than catalog-level data alone suggest.

Enabling technologies (cryogenic, photonic, and materials engineering infrastructure that all three pillars depend on) form a supporting layer of the quantum supply chain that is not yet addressed by any NRT program as a primary curriculum focus.

3. Build student agency into program governance, not just program activities. Programs that give students formal ownership of persistent program elements (quantum device design workshops, course design, recruiting activities, seminar series, and peer mentoring structures) consistently report better outcomes than programs that treat student engagement as supplemental. This is not primarily about student satisfaction; it is about developing the project management, communication, and peer-mentoring skills that industry employers identify as most valued and most often absent in new graduates. Student co-design of courses, as practiced at OSU QuGIP and UCSB InTriQATe, produces courses better calibrated to actual student confusion and gives senior students a form of training unavailable through any other mechanism. These structures also directly address the belonging and identity problem that is particularly acute for engineering and computer science students entering physics-dominated programs: a student who owns something in the program is not a guest in the field. Topic-specific hackathons, i.e., centering a cohort around an authentic open-ended problem for a compressed period, are a particularly scalable implementation of this principle: the UCLA quantum design workshops, which draw roughly 300 participants per year, demonstrate that hackathons can function as learning environments that develop the collaboration, communication, and decision-making capacities that formal coursework rarely reaches, while simultaneously extending the program's community well beyond its funded trainees.

4. Establish structural mechanisms for industrial engagement rather than depending on goodwill. Programs that have sustained meaningful industrial partnerships have done so through mechanisms that make engagement low-friction for companies: fee-for-capstone models (UW AQET), pre-cleared IP templates, and explicitly structured away experiences with fellowship support. FFRDC partnerships deserve more systematic exploration, particularly for sensing and communications programs where the primary applications sit in the defense and national security domain.

5. Design for sustainability from the first year of funding. The five-year NRT funding cycle is sufficient for proof of concept but not for institutional embedding. Programs should, from the outset, identify which program elements are candidates for integration into regular departmental curricula (and plan those courses accordingly), which elements depend on external funding and therefore require alternative revenue mechanisms, and which elements are candidates for portability and adoption by programs beyond the originating institution. The IUSE grant mechanism is one appropriate vehicle for the last category. NRTs could establish alternative sustainability pathways beyond traditional grant mechanisms, including opportunities for patenting, licensing, and downstream technology transfer, and foster an entrepreneurial mindset by engaging trainees within this innovation pipeline. Programs that treat sustainability planning as a year-four activity will find themselves unprepared. Sustainability is not a funding problem alone; it is a recruitment problem and a pipeline problem as well.

6. Develop graduate-level textbooks spanning all three pillars of engineering-oriented QISE. The absence of such textbooks is a material constraint on faculty adoption. The community that produced the IEEE undergraduate roadmap [2] should undertake a parallel effort at the graduate level. The undergraduate textbook Quantum Mechanics for Tomorrow's Engineers [24] addresses quantum devices at the undergraduate level; no comparable graduate text spans quantum computing hardware, quantum sensing, and quantum communication from an engineering perspective. The need is confirmed in practice: educators teaching graduate QISE courses to interdisciplinary cohorts consistently report that Nielsen and Chuang, the de facto standard, is inaccessible to students from engineering and computer science backgrounds without substantial faculty-mediated supplementation, and that the resulting pedagogical shortfall is typically filled with primary literature, a solution that requires expertise and effort that teaching-focused programs cannot reliably sustain [21]. Writing such a text is a multi-year community effort that should begin now, coordinated across the NRT programs and potentially supported by NSF or industry partners as a curriculum development deliverable.

7. Establish shared outcome assessment instruments across NRT programs, and publish existing evaluation findings. This roadmap is itself a practitioner synthesis rather than an evaluation report: we do not provide the pre/post outcome data or psychometric analysis we are recommending, but that makes the prescription more urgent, not less. The assessment gap identified in Section VIII.G cannot be closed by any single program working alone. The community should develop a small set of shared measurement instruments (attitudinal surveys, competency assessments, and alumni tracking metrics) that can be administered across programs with minimal burden, producing cross-institutional data that no individual program can generate independently. The evaluation team at QuaNTRASE (UTK/UGA), which is extending the SURE framework to the NRT context, is a natural convener for this effort. NSF should be encouraged to treat cross-program assessment infrastructure as a fundable deliverable of the NRT portfolio, not just an individual program obligation. A closely related obligation concerns evaluation data that already exists. Most NRT programs have external evaluation reports that document program efficacy, but these data are largely unavailable to the community, locked in NSF reporting systems that are not publicly accessible [16]. Programs should publish their external evaluation findings in the education research literature. NSF should require deposition of evaluation reports in a public repository as a condition of award completion. The cumulative evidence base the field needs cannot be built on data that remains private.

8. Develop structured mechanisms for faculty professional development in QISE. The scaling problem documented in Section VIII.H is, at its core, a faculty pipeline problem: nine of fifteen QISE educators interviewed across a range of institutions identified insufficient QISE faculty expertise as a documented constraint on their program's ability to deliver a full curriculum [23]. No individual institution can solve this, but the NRT community collectively holds the expertise to address it. The community should develop structured pathways for faculty from adjacent fields (condensed matter physics, atomic molecular and optical physics, electrical engineering, computer science, and information theory) to acquire the cross-pillar QISE literacy needed to teach and advise in interdisciplinary programs. Summer institutes, visiting faculty programs that bring researchers from national laboratories and industry into teaching roles, and coordinated NRT faculty exchanges are all vehicles worth developing. NSF and DOE are the natural funders; the NRT network is the natural convener. Without a deliberate investment in faculty development, the programs documented in this roadmap will remain concentrated at the research-intensive institutions where QISE-trained faculty already exist, and the scaling problem will remain structurally unsolved.

## X. CONCLUSION

Graduate training in quantum information science and engineering is at a formative moment. The programs described in this roadmap were built under conditions of significant uncertainty, starting from among the first dedicated graduate QISE programs in the US, including programs at Wisconsin, Chicago, Mines, Duke, and CU Boulder founded in 2019-2020, alongside a rapidly growing international landscape at Waterloo, ETH Zurich, TU Munich, Tsinghua, UTSC, and beyond. It was at first not clear what the quantum workforce would actually need; how interdisciplinary graduate quantum training could work in practice; and whether or not the institutional structures of universities could possibly accommodate the kind of cross-departmental collaboration that QISE requires. The accumulated experience of our eighteen NRT QISE and adjacent programs has substantially reduced some of this uncertainty, even where it has not eliminated it.

Our most important contribution to graduate quantum education is the wealth of examples and strategies for QISE programs that the NRT community has produced. The startup model has emerged as the most widely adopted organizing philosophy, but the programs have shown that it can be instantiated in very different structural forms: standalone degrees and embedded certificates, bootcamp-augmented research programs, multi-institution partnerships, co-advisory models and student-driven governance and hackathons, fee-for-capstone industry engagement, fee-for-facility industry-academia engagement, long-term sustainable training, and FFRDC collaboration, among others. Programs being designed now do not have to start from scratch. They have eighteen programs' worth of operational evidence about what works across what range of institutional conditions.

The catalog of open problems we have provided in Section VIII is our community's honest accounting of what the NRT experience has not resolved. These are (i) the three-pillar imbalance around quantum sensing (and the near-total absence of enabling technologies as an additional curriculum focus); (ii) the missing pedagogical middle for quantum communication; (iii) open quantum systems as invisible substrate; (iv) the bridging-role preparation question; (v) the physical versus virtual training outcomes question; (vi) developing a graduate version of deliberate practice adapted to QISE; (vii) the assessment gap; (viii) the scaling problem; (ix) the formalism entry-point question; (x) sustainability beyond five-year funding; (xi) citizenship restrictions; and (xii) the AI inflection. These are not all the same kind of challenge. Some, such as the three-pillar imbalance, the missing pedagogical middle for quantum communication, open quantum systems pedagogy as engineering substrate, bridging-role preparation, the physical versus virtual training tradeoff, and deliberate practice as a design principle, are tractable at the program level and are already subjects of active curriculum development or pedagogical experimentation. Others, particularly assessment infrastructure, the faculty pipeline, and the formalism entry-point question, whose resolution requires coordination beyond any individual program, require field-wide coordination that no individual institution can initiate alone. A third category comprises problems that reflect external constraints on the community rather than pedagogical questions internal to it: the post-five-year sustainability question every NRT program will eventually face, citizenship restrictions on federal traineeship funding, and the AI inflection the field is navigating in real time. These problems are not program-level solvable and will not be closed by cross-program coordination alone; they require engagement with policy, technology trajectories, and institutional funding structures beyond the NRT portfolio itself.

Our eight concrete recommendations in Section IX are drawn directly from this experience and are addressed to QISE programs being designed or expanded now. These comprise (i) adopting the startup model; (ii) investing in sensing and communication curriculum development as a distinct budget line from trainee support; (iii) building student agency into governance; (iv) creating structural mechanisms for industry engagement rather than depending on goodwill; (v) planning for sustainability from year one; (vi) developing the graduate-level textbooks the community lacks; (vii) establishing shared assessment instruments and publishing existing evaluation data; and (viii) developing structured faculty professional development mechanisms. Every one of these has a proof of concept within the NRT portfolio.

The second quantum revolution will be built by people who are being trained now, in programs that are being designed now. The experiences documented here provide an imperfect but genuine guide to what those programs should look like.

## Acknowledgments

The authors thank Dr. Damon L. Tull, NSF NRT Program Director, for opening the October 2024 satellite meeting and for the NRT program's sustained investment in convergent graduate STEM education. The meeting was supported by the participating NRT programs. This material is based upon work supported by the National Science Foundation under the following NRTs: Mines/SJSU NRT-QL (L. Carr, M. Singh, H. Hurst, E. Khatami, award no. DGE-2125899 and DGE-2125906); UW AQET (K.-M. Fu, S. Mouradian, award no. DGE-2021540); WashU LinQ (K. Murch, S. Hayes, award no. DGE-2152221); OSU QuGIP (J. Gupta, award no. DGE-2244045); Dartmouth NRT (M. Fitzpatrick, E. Fossum, L. Ray, award no. DGE-2125733); UNM QPAQT (V. Acosta, T. Drake, V.E. Babicheva, award no. DGE-2244462); UGA/UTK QuaNTRASE (Y. Abate, G. Siopsis, M. Aydeniz, award no. DGE-2152159); Yale QMSE (S. Ismail-Beigi, award no. DGE-2244310); Rice BRIDGE CQED (J. Kono, A. Alabastri, K. Hazzard, award no. DGE-2346014 ); UCLA AIF-Q (C.W. Wong, award no. DGE-2125924); UCSB InTriQATe (D. Weld, award no. DGE-2152201); UArk NRT-QISE (J. Hu, award no. DGE-2244274); UT Austin Q-CAT (X. Li, award no. DGE-2510551); OKState ATTAQ (Y. Liu, award no. DGE-2510202); MTSU NRT-QISE-AI (W. Ding, award no. DGE-2510724); and FIU NRT Q-STAR (C.-Y. Lai, D. Radu, award no. DGE-2345976). Additional support for quantum sensing curriculum development is based upon work supported by the National Science Foundation under award no. DUE-2417222 (L. Carr, M. Radulaski, B. Marlowe, Mines/UC Davis/MITRE).

## AI and Collaboration Disclosure

This roadmap was drafted collaboratively by the 34-member author team, with AI assistance (Anthropic Claude) used to organize, edit, and refine our original writing session from our workshop. All scientific and educational content, interpretations, findings, and recommendations are those of the author team, including the final draft, and reflect the collective expertise and practitioner knowledge of the NSF quantum NRT community.